\documentclass[notitlepage,twocolumn,prl,nobalancelastpage,amssymb,superscriptaddress,showpacs]{revtex4-1}

\usepackage{Symbols}
\usepackage{Shortcuts}
\usepackage{Text}

\usepackage{epsfig}
\usepackage{subfigure}

\usepackage{pdfpages}

\makeatletter
\AtBeginDocument{\let\LS@rot\@undefined}
\makeatother

\usepackage[colorlinks=true,citecolor=blue,linkcolor=magenta]{hyperref} 

\newcommand{\amp}[2]{\langle#1|#2\rangle}

\def \vk{\boldsymbol{k}}
\def \vq{\boldsymbol{q}}
\def \vr{\boldsymbol{r}}

\def \vs{\boldsymbol{\s}}

\def \bk{\boldsymbol{k}}

\def \cFD{\text{FD}}
\def \pl{\ell}
\def \ps{s}

\begin{document}
\title{Steady states and edge state transport in topological Floquet-Bloch systems}

\date{\today}

\author{Iliya Esin}
\affiliation{\mbox{Physics Department, Technion, 320003 Haifa, Israel}}
\author{Mark S. Rudner}
\affiliation{\mbox{Center for Quantum Devices and Niels Bohr International Academy,}
\mbox{Niels Bohr Institute, University of Copenhagen, 2100 Copenhagen, Denmark}}
\author{Gil Refael}
\affiliation{\mbox{Institute for Quantum Information and Matter, Caltech, Pasadena, CA 91125, USA}}
\author{Netanel H. Lindner}
\affiliation{\mbox{Physics Department, Technion, 320003 Haifa, Israel}}

\begin{abstract}
We study the open system dynamics and steady states of two dimensional Floquet topological insulators: systems in which a topological Floquet-Bloch spectrum is induced by an external periodic drive.  We solve for the bulk and edge state carrier distributions, taking into account energy and momentum relaxation through radiative recombination and electron-phonon interactions, as well as coupling to an external lead. We show that the resulting steady state resembles a topological insulator in the Floquet basis. The particle distribution in the Floquet edge modes exhibits a sharp feature akin to the Fermi level in equilibrium systems, while the bulk hosts a small density of excitations. We discuss two-terminal transport  and describe the regimes where edge-state transport can be observed. Our results show that signatures of the non-trivial topology persist in the non-equilibrium steady state.
\end{abstract}

\maketitle


\paragraph{Introduction ---}
Periodic driving has recently attracted interest as a promising tool for exploring new phases of quantum matter~ \cite{Yao2007,Oka2009,Inoue2010,Kitagawa2010,Jiang2011,Lindner2011,Kitagawa2011,Gu2011,Lindner2013,Delplace2013,Katan2013,Iadecola2013, Kundu2014, Usaj2014, Khemani2016,Harper2017,Else2016b,Potter2016,VonKeyserlingk2016a}. 
Beyond accessing phases resembling those accessible in equilibrium, ``Floquet systems'' also support anomalous, intrinsically non-equilibrium dynamical phases~\cite{Rudner2013, Titum2016, Khemani2016, Harper2017,Else2016b,Potter2016,VonKeyserlingk2016a, Po2016, Else2016,Zhang2017,Choi2017}. 
 Topological properties and spectra of periodically driven systems have been demonstrated in experiments in solid state \cite{Wang2013,Mahmood2016}, cold atoms \cite{Jotzu2014,Aidelsburger2015,Lohse2016,Nakajima2016,Flaschner2016}, and optical systems \cite{Rechtsman2013,Maczewsky2016}.

In this work we focus on Floquet topological insulators (FTIs): systems in which a topological Floquet band structure is induced in a topologically-trivial system by a time-periodic drive~\cite{Lindner2011}. 
Investigating the complex non-equilibrium steady-states that result from the unavoidable coupling to bath degrees of freedom, such as phonons, is essential for understanding the physical properties of Floquet systems~\cite{Dehghani2014,Dehghani2015,Iadecola2015,Iadecola2015a,Seetharam2015,Liu2017}.
In particular, 
when the system is longer than the inelastic mean free path (MFP), transport depends crucially on the interplay between the coupling to the system's leads and to its intrinsic baths. 
We thus seek to characterize these steady states, and to understand their physical manifestations.

\begin{figure}[ht]
  \centering
  \includegraphics[width=0.9\columnwidth]{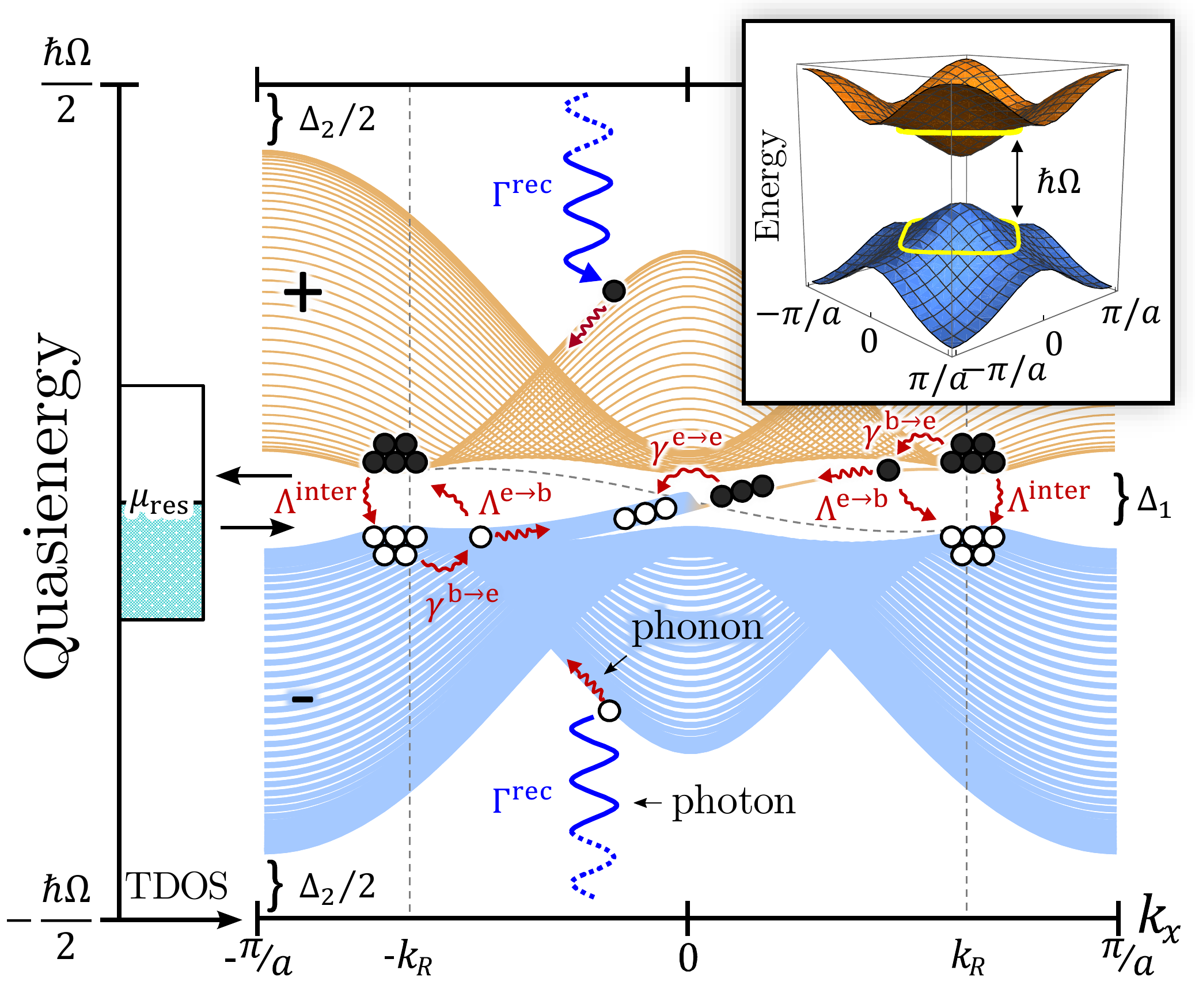}
  \caption{(Color online) Quasienergy spectrum of a 2D Floquet topological insulator in a cylindrical geometry. 
Wavy arrows illustrate processes due to electron-boson interactions, captured phenomenologically in Eq.~\eqref{eq:RateEquation}. Excitations from the lower to the upper bulk Floquet band are mediated by radiative recombination (with rate $\G^{\rm rec}$). Relaxation to the lower Floquet band is mediated by phonons  $(\Lm^{\rm inter})$. Phonons also mediate transitions between the bulk and the edge ($\g^{\rm b\to e}$ and $\Lm^{\rm e\to b}$) and  within the edge $(\g^{\rm e\to e})$. These processes yield an insulatorlike steady state filling of the Floquet bands, with additional electron and hole excitations (filled and empty circles, respectively). The system is coupled to an energy-filtered Fermi reservoir with a narrow effective bandwidth (left).  Inset: The non-driven bulk spectrum of the system. The yellow loops depict the resonance condition of the periodic drive. \label{fig:System}}\vspace{-0.2 in}
\end{figure}

In the present study we consider two-dimensional (2D) systems in which a resonant drive is used to induce a band inversion in the Floquet-Bloch spectrum (see Fig.~\ref{fig:System}). The resulting Floquet bands have non-zero Chern numbers, and in a finite geometry with edges exhibit chiral Floquet edge modes. In this work we will be particularly interested in the steady states of the chiral Floquet edge modes, and their coexistence with the non-equilibrium steady state of the bulk.

In a driven electronic system, the natural intrinsic baths to consider are the phonons of the crystal lattice and the photons of the ambient electromagnetic environment.
  In the system we consider, the role of acoustic phonons is mainly to relax momentum and (quasi)energy, while photon emission associated with particle-hole recombination acts as a primary heating source in the Floquet band picture (similar considerations were applied to one-dimensional systems in~\cite{Seetharam2015}). Due to the edges of the system, the steady state is inhomogeneous, and therefore we analyze the system using a full Floquet-Boltzmann approach~\cite{Genske2015}. To deduce the transport properties of the system, we also consider the effects of a coupling to an external Fermi
reservoir (i.e., a lead).

Below we show that the steady-state, characterized by the populations of Floquet-Bloch states, resembles that of a topological insulator, with an additional non-equilibrium Fermi sea of electrons and holes in the bulk. The chiral Floquet edge states are populated according to a smooth distribution with a well defined Fermi level. In the presence of coupling to an energy-filtered Fermi
reservoir, whose chemical potential lies in the Floquet band gap \cite{Seetharam2015}, we find that: (1) the bulk excitation density is insensitive to variations of the reservoir chemical potential; (2) the Fermi level of the edge states is pinned to the chemical potential of the reservoir. Using these results, we assess the stability of the edge currents and give prospects for measuring edge transport in Floquet topological insulators.


\paragraph{Model of the FTI ---}
We now introduce the model for the driven system. 
We consider a two-band 2D model, described in the absence of driving by the  Hamiltonian
\begin{equation}
\hat\cH_0=\sum_{\vk} \hat c_{\vk\n}\dg \bR{\tb d(\vk)\cdot \vs}_{\n\n'}   \hat c_{\vk\n'},
\label{eq:FreeHamiltonian}
\end{equation}
where $\vs=\bR{\s^x,\s^y,\s^z}$ is the vector of Pauli matrices, and $\hat c_{\vk\n}\dg$ creates an electron with quasimomentum $\vk$ and pseudospin $\n= \bC{\aup,\adn}$.
We take $\tb d(\vk)=(A\sin(ak_x),A\sin(ak_y),M-4B+2B\cos(ak_x)+2B\cos(ak_y))$, such that Eq.~(\ref{eq:FreeHamiltonian}) describes half the degrees of freedom in the BHZ model for time-reversal invariant semiconductor quantum wells~\cite{Dang2014,Bernevig2006,Zitouni2005}.
Here $A,B$ and $M$ are material-dependent parameters, and $a$ is the lattice constant of the crystal.
We assume a trivial semiconductor (with non-inverted band structure), with $M > 0$ and $B<0$.




The semiconductor is periodically driven by an external field with an above-gap frequency $\W$. For simplicity we consider a uniform driving field of amplitude $V_0$ that couples to electrons through $\s^z$ \footnote[10]{More realistic time-dependent electromagnetic fields can be incorporated in this model, see \cite{Lindner2011}}, modeled by the time-dependent Hamiltonian
\begin{equation}
\hat\cH_V(t)=\half V_0\cos(\W t)\sum_{\vk} \hat c_{\vk\n}\dg\s^z_{\n\n'}\hat c_{\vk\n'}.
\label{eq:DrivenHamiltonian}
\end{equation}

Below we work in the basis of Floquet-Bloch eigenstates of the time-periodic single particle Hamiltonian  $\hat \cH_0+\hat \cH_V(t)=\sum_{\vk}\hat c_{\vk\n}\dg \bS{H(t)}_{\vk,\n\n'} \hat c_{\vk\n'}$. The Floquet eigenstates satisfy $\bR{i\hb\frac{\dpa }{\dpa t}-H(t)}\ket{\y(t)}=0$, with $\ket{\y(t)}=e^{-i\ve t/\hb}\ket{\phi(t)}$. Here  $\ket{\f(t)} = \ket{\f(t + T)}$ is periodic with period $T = 2\pi/\Omega$,  and $\ve$ is the quasienergy. Throughout, we use the convention $-\hb\Omega/2\leq\ve<\hb\Omega/2$.





 The driving field yields resonant transitions between the valence and conduction bands along a closed curve in momentum space, see Fig.~\ref{fig:System} (inset). A gap of magnitude $\Delta_1 \propto |V_0|$ opens at quasienergy $\ve=0$,  yielding two separate quasienergy bands.
 The driving field leads to an effective band inversion of the Floquet bands with respect to the original non-driven band structure. An important consequence of this band inversion is the appearance of chiral edge states in the gap at $\ve=0$ for a system in a finite geometry with edges \cite{Lindner2011}. We restrict $\hb\Omega>|M-8B|$, such that there is only a single-photon resonance.

We label the bulk Floquet states by the quasimomentum $\vk$ and a Floquet band index $\a=\pm$ (distinct from the band index of the non-driven system): $\ket{\y_{\vk\a}(t)}=e^{-i\ve_{\a}(\vk) t/\hb}\sum_{m}e^{im\W t}\ket{\f^m_{\vk\a}}$ \cite{Sambe1973,Shirley1965}. We refer to the Floquet bands with quasienergies $0<\ve<\hb\W/2$  and $-\hb\W/2<\ve<0$ as the upper $(+)$ and lower $(-)$ Floquet bands, respectively, see Fig.~\ref{fig:System}.

\begin{figure}
  \centering
  \includegraphics[width=8.6cm]{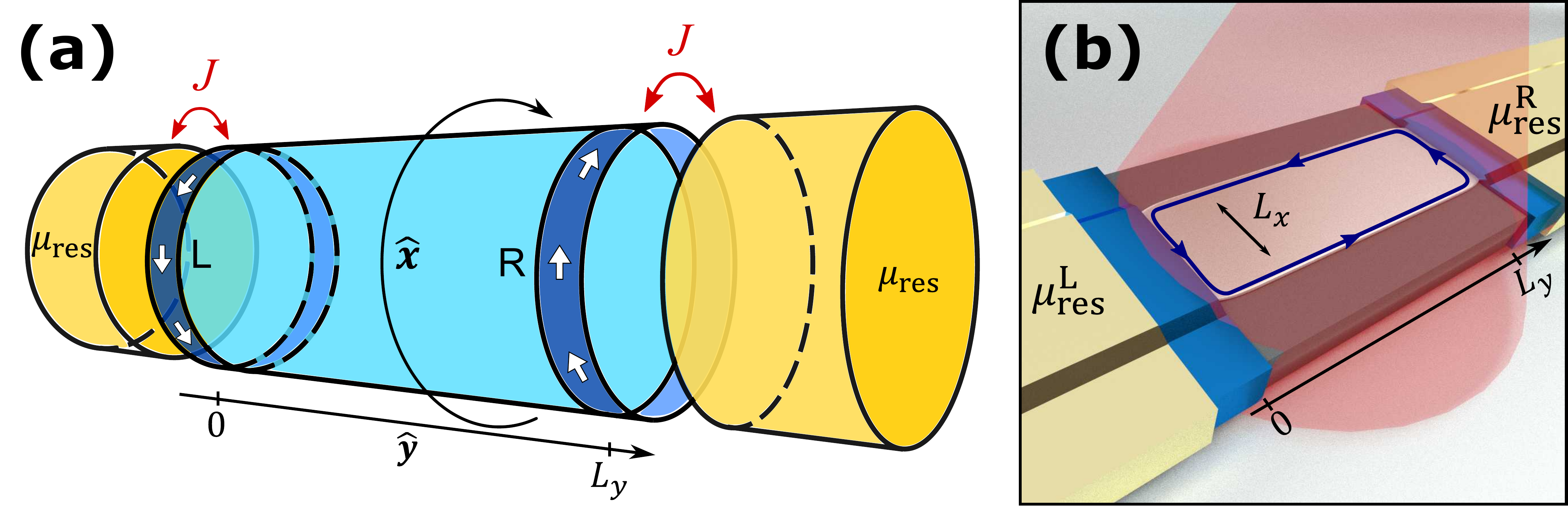}
  \caption{(Color online) (a) Schematic drawing of the system coupled to leads in the cylinder geometry. Dark blue rings indicate the right (R) and left (L) edge states. The energy filtered leads are set to have equal chemical potentials, $\m_{\rm res}$, coupling strength to the system $J$, and density of states. (b) Two-terminal transport geometry. Contacts (yellow) are connected to a periodically driven semiconductor (white) through an energy filter (blue). \label{fig:Geometry}}
\end{figure}

In the following, we will consider a system with periodic boundary conditions in the $x$ direction, and open boundary conditions in the $y$ direction. As seen in Fig.~\ref{fig:System}, in this geometry the edge states exist for quasimomentum $k_x$ in the interval $-k_R\lesssim k_x\lesssim k_R$, where $k_R$ is the maximal value of $k_x$ for which the driving field is resonant. We denote the Floquet edges states as $\ket{\chi_{k_x\be}(t)}$, where the label $\be$ corresponds to the  left (L) and right (R) edges (at $y=0$ and $y=L_y$), for which $\dpa \ve/\dpa k_x$ is negative and positive, respectively, see Fig.~\ref{fig:Geometry}a.

\paragraph{Coupling to a bosonic heat bath ---}
The open, driven system evolves to a steady state, governed by its coupling to one or more heat baths (taken to be at zero temperature). We first focus on the bosonic bath, and consider the roles of acoustic phonons and photons (associated with radiative recombination).   Using the label $\lm=\pl,\ps$ to denote the photon (light) and acoustic phonon (sound) modes, we describe the dynamics of each mode by the Hamiltonian
\begin{equation}
\hat\cH_\lm=\sum_{\vq}\hb v_\lm \abs{\vq}\bR{\hat b\dg_{\lm,\vq} \hat b_{\lm,\vq}+\half}.
\end{equation}
Here $\hat b_{\lm,\vq}\dg$ are creation operators of  $\lm$-bosons.
The velocity $v_\lm$ is taken to be constant and isotropic for each mode.
While the electronic degrees of freedom are confined to a 2D plane, we take the bosonic bath modes to live in three dimensions; for simplicity we consider a single polarization mode for each boson type. For the (finite bandwidth) acoustic phonon bath, we take a linear dispersion up to a Debye frequency, $\omega_D$ \footnote[40]{In this work we use simple models for the acoustic phonons and the electromagnetic environment, and their couplings to the system.  More detailed modeling of these baths would not qualitatively change our results.}.  

Inspired by the physics of semiconductor quantum wells, we assume that emission of a photon is accompanied by a pseudo-spin flip (corresponding to a change of one unit of electronic angular momentum). Furthermore, we take the interaction with acoustic phonons to conserve the pseudospin index, as acoustic phonons have suppressed matrix elements between different atomic orbitals. The Hamiltonian describing local interactions between electrons and $\lm$-bosons thus reads:
\begin{equation}
\begin{split}
\hat\cH_{e-\lm}=&\sum_{\vr} \hat c_{\vr\n}\dg\bS{\h_{\lm,\n\n'}\dg \hat b\dg_{\lm,\vr}+\h_{\lm,\n\n'}\hat b_{\lm,\vr}}\hat c_{\vr\n'},
\label{eq:HamiltonianElectronBoson}
\end{split}
\end{equation}
where $\h_{\ps}=g_\ps\I$ for electron-phonon coupling, and $\h_{\ell}=g_{\pl}\s^+$ for electron-photon coupling. The quantities $g_\pl$ and $g_\ps$ denote the associated coupling strengths.
In Eq.~(\ref{eq:HamiltonianElectronBoson}), the coordinate $\vr$ is confined to the 2D plane.

In closing this section defining the model, we note that the full Hamiltonian possesses particle-hole and inversion symmetry at all $t$. The system's Floquet spectrum  and the kinetic equations derived below exhibit corresponding symmetries. However, our qualitative conclusions do not depend on these symmetries.


\paragraph{Phenomenological model for the steady state ---}
Before diving into the full kinetic equation, we first characterize
the steady states using a simplified phenomenological model, which takes into account the most significant contributions to the population kinetics in the system (see Fig.~\ref{fig:System}).  In the following discussion, we restrict our attention to a half-filled system.

Generically, the population kinetics in a driven system differs from that of a system in thermal equilibrium, due to scattering processes in which the total quasienergies of the incoming and outgoing modes differ by integer multiples of $\hbar\Omega$. As a starting point, we first consider a system in which the sums of quasienergies of the incoming modes and outgoing modes are strictly equal in all scattering processes (which requires the system-bath coupling to obey special conditions~\cite{Galitskii1970,Liu2015,Shirai2015,Shirai2016,Iwahori2016}). In this situation, the steady state of the driven system is simply given by a Fermi-Dirac distribution in terms of the Floquet bands, with the ordering of quasienergies (i.e., choice of Floquet-Brillouin zone) as used in Fig.~\ref{fig:System}. The temperature of the distribution is that of the phonon bath. For a half-filled system, we obtain an ideal FTI: when the bath is at zero temperature, the lower (upper) Floquet band is filled (empty), and the edge state is filled up to the Fermi level at $\varepsilon=0$ (corresponding to $k_x=0$).


Our goal is to obtain the steady state of the system in the presence of \ti{all} scattering processes, including those where the total quasienergy changes by a multiple of $\hbar\Omega$. These ``Floquet-Umklapp'' processes create excitations from the lower to the upper Floquet band, and thereby act as a source of ``heating'' in the Floquet basis. We characterize the steady state in the bulk by the density of excited electrons in the ``upper'' (+) bulk Floquet band, $n_{\rm b}=\int \frac{d^2\vk}{(2\p)^2} \av{\hat \y\dg_{\vk+}(t)\hat \y_{\vk+}(t)}$. At each edge the steady state is characterized by the density of excited particles above the Fermi level of the ideal FTI ($\ve = 0$). For the right edge, this density is given by $n_{\rm e}=\int_{0}^{k_R} \frac{dk_x}{2\p} \av{\hat\ch\dg_{k_xR}(t)\hat \ch_{k_xR}(t)}$. The operators  $\hat \y\dg_{\vk\a}(t)$ and $\hat \chi\dg_{k_x\be}(t)$ create electrons in the bulk and edge Floquet states $\ket{\y_{\vk\a}(t)}$ and $\ket{\chi_{k_x\be}(t)}$, respectively \footnote[50]{The operators  $\hat \y\dg_{\vk\a}(t)$ and $\hat \chi\dg_{k_x\be}(t)$ obey the anticommutation relations  $\{\hat \y\dg_{\vk\a}(t),\hat \y_{\vk'\a'}(t)\}=\delta_{\vk \vk'}\delta_{\a\a'}$ and $\{ \hat \chi\dg_{k_x\be}(t),\hat \chi_{k_x'\be'}(t)\}=\delta_{k_x k_x'}\delta_{\beta\beta'}$.}. The distributions of electrons in states with $\varepsilon>0$ and of holes in states with $\varepsilon<0$ are related by particle hole symmetry (see below). Additionally, the distributions in the right and left edge states are related by inversion symmetry.


For a semiconductor with a sufficiently large band gap, such that $M\gg\hbar\omega_{\rm D}$, Floquet-Umklapp processes resulting from phonon scattering are suppressed as $[V_0/(\hb\Omega)] ^4$ \cite{Seetharam2015}. For simplicity, in our analysis we will assume that all Floquet-Umklapp process are due to radiative recombination.  Since this process involves emission of a photon, it predominately contributes when the characters of the initial and final states correspond to the conduction and valence bands of the undriven system, respectively (recall that the electron-photon coupling is off-diagonal in pseudospin). Close to the ideal FTI steady state, $\vk$-modes in the lower Floquet band with momenta inside the
resonance curve are filled, and have a conduction band character, while those of the upper band are empty and have valence band character. Radiative recombination between these states leads to a source term for particles in the upper Floquet band, $\dot{n}_{\rm b}=\Gamma^{\rm rec}$ (see Fig.~\ref{fig:System}), with rate $\Gamma^{\rm rec}$ approximately independent of the excitation density for small deviations from the ideal FTI state.

Once excited to the upper Floquet band, electrons quickly relax to the band minima due to scattering by phonons.
Near the Floquet band minima (around the resonance curve), the Floquet states are hybridized superpositions of valence and conduction band states.
This hybridization allows phonons to scatter electrons from these minima to empty states near the maxima of the lower Floquet band.
Consider the rate of such phonon-assisted ``recombination'' of Floquet-band carriers.
 During such a process, an electron in the upper band must find a hole in the lower band. 
The resulting rate is thus proportional to the density of electrons times that of the holes (which are equal at half filling): 
$\dot{n}_{\yb} \approx -\Lambda^{\rm inter} n_\yb^2$.


Next, we account for processes which scatter particles between bulk and edge states.
Such bulk-edge scattering processes are predominantly phonon-assisted (the rates for photon-assisted bulk-edge scattering are suppressed by a small density of states).
Assuming a small population of excited electrons (with $\ve>0$) on the edge, $n_{\rm e}\ll1/a$, bulk-to-edge processes predominantly take excited electrons in the upper Floquet band to the nearly empty $k$-space region of the edge states (with $k_x>0$, for the right edge). In contrast, edge-to-bulk processes require that the scattered edge electron finds an empty bulk state (i.e., a hole) in the lower Floquet band (see Fig.~\ref{fig:System}). The corresponding rate is thus proportional to both the densities of excitations on the edge and in the bulk. We therefore estimate the contribution of bulk-edge processes to $\dot n_{\rm e}$ as $\dot n_{\rm e}=\g^{\rm b\to e}n_{\rm b}-\Lm^{\rm e\to b}n_{\rm b} n_{\rm e}$. The parameters $\g^{\rm b\to e}$ and $\Lambda^{\rm e\to b}$  encode the rates of bulk-to-edge and edge-to-bulk scattering processes, respectively.


Last, we account for phonon-assisted scattering of particles within the edge. 
At low phonon temperatures, such processes predominately decrease the quasienergy of the electrons, and thus tend to decrease the density of excited particles on the edge. The requirement that an excited edge-electron finds an edge-hole gives $\dot n_{\rm e}=\gamma^{\rm e\to e} n_{\rm e}^2$.



Summing up the processes above, we arrive at the rate equations for the bulk and edge excitation densities:
\begin{subequations}
\begin{eqnarray}
\!\!\!\!&&\dot n_{\rm b}= \G^{\rm rec}- \Lm^{\rm inter}n_{\rm b}^2-\frac{2}{L_y}\bR{\g^{\rm b\to e} n_{\rm b}-\Lm^{\rm e\to b}n_{\rm b} n_{\rm e}}
\label{eq: RateEquation a}\\
\!\!\!\!&&\dot n_{\rm e}=\g^{\rm b\to e}n_{\rm b}-\Lm^{\rm e\to b} n_{\rm b} n_{\rm e} -\g^{\rm e\to e}n_{\rm e}^2.
\label{eq: RateEquation b}
\end{eqnarray}
\label{eq:RateEquation}
\end{subequations}
The steady state solution for the above equations is obtained for $\dot n_{\rm b}=\dot n_{\rm e}=0$.

In the thermodynamic limit, the rate parameters in Eq.~(\ref{eq:RateEquation}) become independent of system size \cite{SeeSM}.
Note that in Eq.~(\ref{eq: RateEquation a}), the source term for the 2D density $n_{\rm b}$ due to coupling to the 1D edge is multiplied by a factor of $1/L_y$.
Thus for $L_y \to \infty$, Eq.~(\ref{eq: RateEquation a}) yields a bulk excitation density $n_{\rm b}$ which is independent of $n_{\rm e}$, and scales as
\begin{equation}
n_{\rm b} \eqa\ka^\half, \quad \ka=\G^{\rm rec}/\Lm^{\rm inter}.
\label{eq:ModelResults}
\end{equation}
As expected, the bulk excitation density is unaffected by the presence of the edge. The dimensionless parameter $\ka a^4$  captures the competition between ``heating'' (Floquet-Umklapp) and ``cooling'' processes in the bulk.

The rates controlling the excitation density on the edge in Eq.~(\ref{eq: RateEquation b}) are predominantly due to phonon-assisted scattering.
Therefore their ratios do not scale with $\kappa$.  For sufficiently small $\kappa$, we reach $\frac{\g^{\rm e\to e}\g^{\rm b\to e}}{(\Lm^{\rm e\to b})^2}\gg n_\yb $. In this limit, the second term in Eq.~(\ref{eq: RateEquation b}) can be omitted and we find for the steady state: 
\begin{equation}
 n_{\rm e} \eqa\bR{\g^{\rm b\to e}/\g^{\rm e\to e}}^{\half}\ka^\quat,
\label{eq:ModelResults2}
\end{equation}
where the ratio $\g^{\rm b\to e}/\g^{\rm e\to e}$ is independent of $\ka$.

The bulk excitation density $n_\yb$ estimated in Eq.~(\ref{eq:BulkPopulation}) represents a spatial average over the full system.
In a more detailed picture, we expect the excitation density to be inhomogeneous, deviating from the bulk value estimated in Eq.~(\ref{eq:BulkPopulation}) near the edges.
We investigate the spatial dependence of $n_{\rm b}$ below.



\paragraph{Microscopic analysis of the steady state ---}
We now turn to a more microscopic treatment, and characterize the steady state using a Floquet-Boltzmann equation approach. We focus on the regime where the MFP is larger than the characteristic wavelength of electrons. We characterize the steady state in the bulk in terms of a phase space distribution function $f^\yb_{\vk\a}(\vr;t)$. Due to the translational invariance of the cylinder, we assume that the phase space distribution is independent of $x$.  Therefore we define:
\begin{equation}
\! f^\yb_{\vk\a}(y;t)=\frac{L_y}{\p}\int dk_y' e^{2ik_y' y}\av{\hat \y_{\vk+k_y'\hat{\boldsymbol{y}}\a}\dg(t) \hat \y_{\vk- k_y'\hat{\boldsymbol{y}}\a}(t)}\label{eq:BulkPopulation}.
\end{equation}
Note that $\int\!\!\frac{d^2\vk}{(2\pi)^2}f^\yb_{\vk\a}(y;t)$ gives the density of electrons in band $\alpha$ at position $y$ (for any $x$), at time $t$.  
A dependence on $y$ is expected due to the edges at $y=0,L_y$~\footnote[12]{Off-diagonal correlations between states separated with large gaps, on the scale of the scattering rates vanish \cite{Hone2009,Seetharam2015}.}.  The distributions within the one-dimensional edge states are defined as
$f^\ye_{k_x\be}(t)=\av{\hat \ch_{k_x \be}\dg(t) \hat \ch_{k_x \be}(t)}$.


Next, we study the steady-state behaviour of $f^\yb_{\vk\a}(y)$.
The physics on length scales larger than the MFP is  described by the Floquet-Boltzmann equation~\cite{Genske2015},
\begin{equation}
\dpa_t f^\yb_{\vk\a}+v_{y,\a}(\vk) \dpa_{y} f^\yb_{\vk\a}=\cI^{\yb\yb}_{\vk\a}+\cI_{\vk \a}^{\yb\yR}+\cI_{\vk \a}^{\yb\yL}.
\label{eq:WignerBoltzmannKineticEquation}
\end{equation}
Here $v_{y,\a}(\vk)=\hb\inv\dpa_{k_y} \ve_\a(\vk)$ is the Floquet band group velocity in the $y$ direction, and the collision integrals $\cI^{\yb\yb}_{\vk\a}$, $\cI^{\yb\yR}_{\vk\a}$, and $\cI^{\yb\yL}_{\vk\a}$  describe bulk-bulk, bulk-right-edge and bulk-left-edge scattering processes, respectively. For brevity, in Eq.~\eqref{eq:WignerBoltzmannKineticEquation} we used $f^\yb_{\vk\a}\equiv f^\yb_{\vk\a}(y;t)$; likewise, we suppressed the dependence of the collision integrals on $y$ and $t$.
The Boltzmann equation for the edges has a similar structure, namely, $\dpa_t f^\ye_{k_x \be}=\cI^{\ye\ye}_{k_x\be}+\cI_{k_x \be}^{\ye\yb}$.

In explicit form, the collision integral for bulk-to-bulk scattering processes is given by
\begin{equation}
\cI^{\yb\yb}_{\vk\a}\!=\!\sum_{\vk'\a'}\bS{W_{\vk'\a'}^{\vk\a}f^\yb_{\vk'\a'}(1-f^\yb_{\vk\a})-W_{\vk\a}^{\vk'\a'}f^\yb_{\vk\a}(1-f^\yb_{\vk'\a'})},
\label{eq:CollisionIntegral}
\end{equation}
where $W_{\vk'\a'}^{\vk\a}$ is the total scattering rate from $(\vk,\a)$ to $(\vk',\a')$. The rates $W_{\vk'\a'}^{\vk\a}$ in Eq.~\eqref{eq:CollisionIntegral} are $y$-independent, and therefore any $y$ dependence of $\cI^{\yb\yb}_{\vk\a}$ arises through the distributions $f^\yb_{\vk\a}(y;t)$. In contrast, for the bulk-edge collision integrals $\cI_{\vk \a}^{\yb\yR}$ and $\cI_{\vk \a}^{\yb\yL}$ the corresponding rates themselves are only significant for values of $y$ near the edges, due to the spatial profile of the edge states.  The full expressions for all the collision integrals can be found in the Supplementary Material~\cite{SeeSM}.

The rate $W_{\vk\a}^{\vk'\a'}$ in Eq.~\eqref{eq:CollisionIntegral} can be written as a sum of phonon ($s$) and photon ($\ell$) assisted scattering rates, $W_{\vk\a}^{\vk'\a'}=W_{\pl,\vk\a}^{\vk'\a'}+W_{\ps,\vk\a}^{\vk'\a'}$, given by
\begin{equation}
\begin{split}
W_{\lm,\vk\a}^{\vk'\a'}=\frac{2\p}{\hb}&\sum_n\abs{\sum_m\braoket{\f^m_{\vk\a}}{\h_{\lm}}{\f^{m-n}_{\vk'\a'}}}^2\times \\ \times&\ro_{\lm}\bR{\ve_\a(\vk)-\ve_{\a'}(\vk')+n\hb\W,\vk-\vk'}.
\end{split}
\label{eq:Wmatrix}
\end{equation}
The DOS of $\lm$-bosons is given by 
$\ro_{\lm}(\ve,\vq)=\frac{a^2}{L_xL_y}\frac{a\ve\Q(\ve-\hb v_\lm\abs{\vq})}{\p \hb v_\lm\sqrt{\ve^2-\hb^2 v_\lm^2\abs{\vq}^2}}$, where (as above) $\lambda = \{s, \ell\}$.
For relatively low energy emission processes [e.g., relaxation across the Floquet gap, contributing to $\Lambda^{\rm inter}$ in Eq.~(\ref{eq: RateEquation b})], the photon DOS is suppressed relative to the phonon DOS by $v_s/v_\ell$ and phonon-emission dominates.
For high energy transfers, the DOS of phonons  vanishes when $\ve$ is above the Debye energy, $\hb \w_{D}$.
In this work we fix $\omega_D$ within the range $\D_{1}<\hb \w_D<\D_{2}$, ensuring 
Floquet-Umklapp processes induced by phonon scattering are fully suppressed.
Here $\Delta_1$ and $\Delta_2$ are the gaps centered at $\varepsilon = 0$ and $\varepsilon = \hb\Omega/2$, respectively, see Fig.~\ref{fig:System}.

Within this formalism, we can estimate the phenomenological rates in the effective model, Eq.~\eqref{eq:RateEquation}, using microscopic parameters (for full details see~\cite{SeeSM}).
We denote by $\mathcal{W}_{\bk}^{\rm rec}=\bR{\frac{L_xL_y}{4\p}\frac{\W^2}{v_\pl^2}}W_{\pl,\vk-}^{\vk+}$ the recombination rate for particles initially in the lower Floquet band.
This rate is significant within the resonance curve, where the Floquet bands are inverted and the characters of the initial and the final states correspond to the conduction and valence bands of the non-driven system, respectively. Thus the source term for the bulk excitation density is $\G^{\rm rec}\eqa\int \frac{d^2\vk}{(2\p)^2} \mathcal{W}^{\rm rec}_{\bk} \eqv \frac{A_{\cR}}{(2\p)^2}\overline{\cW}^{\rm rec}$, where $A_{\cR}$ is the momentum-space area inside the resonance curve.
We estimate the parameter $\Lambda^{\rm inter}$  characterizing phonon-assisted relaxation between Floquet bands as $\Lambda^{\rm inter}\eqa L_xL_y \overline{\mathcal{W}}^{\rm inter}$, where   $\overline{\mathcal{W}}^{\rm inter}=W_{\ps,\vk_R +}^{\vk_R -}$ is an average relaxation rate of a particle in the active region around the minimum of the upper Floquet band. With these definitions, we obtain an approximate expression for $\ka$ 
in Eq.~(\ref{eq:ModelResults}): 
$\ka \approx \frac{A_{\cR}\W^2 v_\ps g_\pl^2}{8 \p^3 v_\pl^3 g_\ps^2}$. The parameters $\gamma^{\rm b\to e}$, $\Lambda^{\rm e\to b}$, and $\gamma^{\rm e\to e}$ can be estimated using the bulk-to-edge and edge-to-edge scattering rates in the same manner.

\begin{figure}
  \centering
  \includegraphics[width=8.6cm]{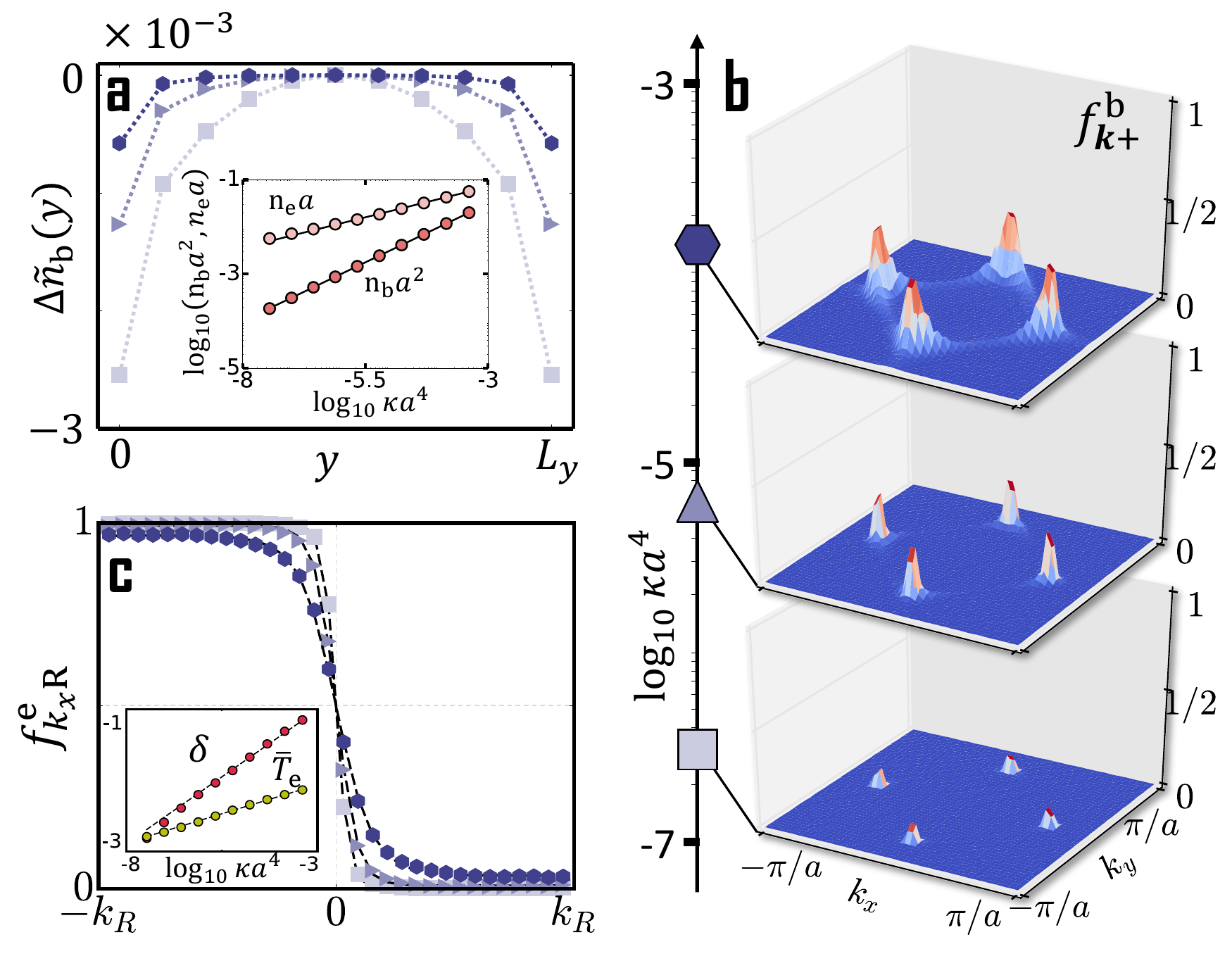}\\
  \caption[0]{(Color online) Steady state of electrons in a half-filled system.
The simulation was performed using a discretization with a $50\times50$ grid in momentum space, and $11$ strips in the $y$ direction (of width comparable to the healing length $\xi$, see text)~\cite{SeeSM}.
(a) Change in spatial dependence of the excitation density in the bulk bands, normalized by the excitation density deep in the bulk, $\D\tilde{n}_\yb(y)=\bR{n_\yb(y)-n_\yb^0}/n^0_{\rm b}$, with $n^0_{\rm b} = n_\yb(L_y/2)$, for three values of $\kappa$.
The color code indicating the values of $\ka$ appears to the right. 
Inset: Scaling of the bulk ($n^0_{\rm b} a^2$) and edge ($n_{\rm e} a$) excitation densities with $\ka a^4$, and the fits $n_\ye a\sim (\ka a^4)^{0.26}$, $n^0_\yb a^4\sim (\ka a^4)^{0.5}$ (black lines).
(b) Distribution of particles in the upper Floquet band ($f^\yb_{\vk+}$) far away from the edges, for different values of $\ka$.
(c) Carrier distribution of the right edge ($f_{k_x\yR}^\ye$) for the same values of $\ka$ as in (a) and (b), and the non-linear least-squares fit to the quasi Fermi-Dirac distribution (dashed lines).
Inset: Effective temperature of the edge, $\bar T_\ye=k_\yB T_\ye/\hb\W$, and the $\dl$-parameter of the quasi Fermi-Dirac function, {\it vs.}~$\ka a^4$.
Dashed lines represent the fits $\bar T_\ye\sim (\ka a^4)^{0.19}$, and $\dl\sim (\ka a^4)^{0.45}$. \label{fig:Results}}
\end{figure}

\paragraph{Numerical simulations ---}
We now numerically solve Eq.~\eqref{eq:WignerBoltzmannKineticEquation} in the steady state, taking $\dot f_{\vk\a}=0$.
We consider the system at half-filling. Figure \ref{fig:Results}a shows the spatial dependence of the bulk excitation density, $n_\yb(y)=\int d^2\vk f_{\vk+}(y)$, for three values of $\kappa$.
Away from the edges, the density reaches a position-independent ``bulk'' value, $n^0_\yb$.
The dependence of $n^0_\yb$ on $\kappa a^4$ is shown in the inset of Fig~\ref{fig:Results}a, and agrees well with our estimate in  Eq.~\eqref{eq:ModelResults}.

The spatial dependence of $n_{\rm b}(y)$ can be accounted for by generalizing Eq.~(\ref{eq: RateEquation a}) to a reaction-diffusion  equation ~\cite{Seetharam2015, SeeSM}.
From this picture we extract the ``healing length'' $\xi$ over which the excitation density relaxes to the bulk value $n^0_{\rm b}$: $\xi \approx \sqrt{D n^0_{\rm b}/(2\Gamma_{\rm rec})}$, where $D$ is the diffusion constant.
Taking $D \approx \bar{v}^2 \tau$, where $\bar{v}$ is a typical velocity of the excited carriers in the steady state and $\tau$ is the scattering time (due to phonons), we find good agreement with the length scales exhibited in our numerical results~\cite{SeeSM}.

%

 Figure~\ref{fig:Results}b shows steady state distributions of the bulk far away from the edges, for three different values of $\ka a^4$. The steady state distribution of the upper band is well described by a Floquet-Fermi-Dirac distribution (a Fermi-Dirac distribution in terms of the quasienergy spectrum), with an effective temperature and chemical potential obtained as fitting parameters. The distribution of the lower band is related by particle-hole symmetry, $f_{\vk,-}^\yb= 1-f_{-\vk,+}^\yb$. The chemical potential describing the distribution in the upper band does not lie in the middle of the gap. Therefore, to describe the distribution of the system, we must use two separate Fermi-Dirac distributions, with distinct chemical potentials, for the upper and lower Floquet bands (for full analysis of the fit to the Floquet-Fermi-Dirac distribution, see~\cite{SeeSM}). Analogous distributions were found for a 1D system in Ref.~\cite{Seetharam2015}. In the absence of photon-assisted recombination (i.e., when $\ka a^4\to 0$), the steady state converges to a global zero-temperature Gibbs state over the Floquet spectrum \cite{Galitskii1970, Liu2015, Shirai2015}.


The steady state distribution of the particles along the right edge is shown in Fig.~\ref{fig:Results}c. The distribution of the left edge is related by inversion symmetry, $f_{k_x\yL}^\ye= f_{-k_x\yR}^\ye$. We observe that the excitations in the edge states predominantly accumulate near $k_x\sim 0$.
The shape of the distribution is  approximated to a good accuracy by a ``quasi Fermi-Dirac distribution,'' defined as $f_{\rm QFD}(\ve)=(1-\dl)f_{\rm FD}(\ve,T_\ye)+\half\dl$.
Here $f_{\cFD}(\ve,T_\ye)$ is the conventional Fermi function, which we scale by a contrast factor ($0<\dl<1$) to create $f_{\rm QFD}$.
The form of the function $f_{\rm QFD}$ dictates that the effective temperature $T_\ye$ is approximately proportional to the excitation density on the edge, $n_\ye$. The $\dl$-parameter describes a small density of particles (holes), uniformly spread along the $k_x>0$ ($k_x<0$) part of the edge mode. The electron and hole ``pockets'' at the extrema of the bulk Floquet bands  provide the source for this excess density.
Thus, we expect $\dl$  to exhibit a similar scaling with $\kappa$ as the density of bulk electrons $n_\yb$.
The dependence of  $n_{\rm e} a$, and of the fitted parameters $T_\ye$ and $\dl$ on $\ka a^4$ are shown in Fig.~\ref{fig:Results}a (inset) and Fig~\ref{fig:Results}c. The results of our simulations are in a good agreement with Eqs.~\eqref{eq:ModelResults} and (\ref{eq:ModelResults2}) and the scaling arguments above.

\paragraph{Coupling to a Fermi reservoir ---}

Can the topological properties of FTIs be identified by transport measurements? To study this question, we couple the system to Fermi 
reservoirs at the two edges, $y = 0$ and $y = L_y$, see Fig.~\ref{fig:Geometry}a. 
The Hamiltonian describing the right reservoir and its coupling to the system reads
\begin{equation}
\hat \cH^{\yR}_{\rm res}=\sum_{l p} \bR{J_{lp}\hat d_l\dg \hat c_{p}+{\rm h.c.}}+\sum_l (\mathcal{E}_{l}-\m_{\rm res}) \hat d_l\dg \hat d_l.
\label{eq:HamiltonianFermiReservoir}
\end{equation}
Here we have introduced a super-index $p$ labeling system operators, Fourier transformed with respect to $x$:  $p=\{k_x,y,\nu\}$. Furthermore,
 $\hat d_l\dg$ is the creation operator for an electron 
in mode $\ket{l}$ of the right reservoir. For simplicity, we choose a system-lead coupling that does not introduce a preferred direction in pseudo-spin space. This is accomplished by taking two degenerate sets of modes, labeled by $l=\{k_x,\mathcal{E}_l,\nu\}$, where $\mathcal{E}_l$ is the mode's energy (which is independent of $\nu=\{\uparrow,\downarrow\}$). The left reservoir and its coupling to the system are described in an analogous manner.
We first consider the left and right reservoirs to have a common chemical potential, $\m_{\rm res}$.

In general, the values of the couplings $J_{lp}$ depend on the precise forms of the reservoir states $\ket{l}$, and the details of the lead-system coupling.
We  take the couplings to be uniform in the $\hat{\boldsymbol{x}}$ direction; for the right lead, we specify $J_{lp} = J \delta_{yL_y}\delta_{\nu(p)\nu(l)}\delta_{k_x(p)k_x(l)}$. For the left lead we replace $\delta_{yL_y}$ with $ \delta_{y0}$. (We do not expect our results to change qualitatively for other generic forms of the reservoirs and the couplings.)

In the following we will consider the effect of the leads when $\m_{\rm res}$ is placed within the Floquet gap. Note that a Floquet state of the system with quasienergy $\ve$ is coupled to reservoir states in a wide range of energies $\mathcal{E}_l=\ve+n\hb \W$ via the harmonics   $\ket{\ch_{k_x\be}^n}$ (or $\ket{\f_{\vk\a}^n}$). As a result, if the reservoir's density of states has a wide bandwidth, electrons 
occupying lead states below the Fermi level can tunnel into the upper Floquet band of the system. These processes (and similar processes for holes) increase the number of excited particles (holes) in the upper (lower) Floquet band, leading to deviations from the ideal Floquet insulator state. To avoid this deleterious effect, we couple the Fermi reservoir through a narrow band of ``filter'' states~\cite{Seetharam2015}, which effectively limits the density of states of the Fermi reservoir.  In our simulation, we take the reservoirs to have a box-shaped DOS of width $w$, aligned symmetrically around the center of a single Floquet zone, see Fig.~\ref{fig:System}.

The introduction of the system-lead coupling, $\hat \cH^{R(L)}_{\rm res}$, 
adds additional collision integrals to the Boltzmann equations for the bulk and edge distributions.
The collision integral describing scattering between the right reservoir and the right edge state 
is given by
\begin{equation}
  \mathcal{I}^{\ye,{\rm res}}_{k_x \yR} = \sum_n \cJ_{k_x\yR}^n\bS{f_{\cFD}\bR{\ve^n_\yR(k_x)-\m_{\rm res}}-f^\ye_{k_x\yR}}.
\end{equation}
Here $\cJ_{k_x\yR}^n=\frac{2\p}{\hb}|J|^2\sum_{l,\nu}\abs{\amp{k_x,L_y, \nu}{\chi_{k_x ;\yR}^n}}^2\dl(\ve^n_\yR(k_x)-\mathcal{E}_l)$, where $|k_x,L_y,\nu\rangle$ is the state created by $c^\dagger_{k_x,y=L_y,\nu}$ and  $\ve_\yR^n(k_x)=\ve_\yR(k_x)+n\hb \W$; $\ve_\yR(k_x)$ is the quasienergy of the right edge state, with quasimomentum $k_x$.
The values of $\mathcal{E}_l$ are limited to the range within the filter window.
An identical expression holds for the left edge state, with $\yR \to \yL$.
In addition, Eq.~(\ref{eq:WignerBoltzmannKineticEquation}) contains a collision integral $\mathcal{I}^{\yb,{\rm res}}_{\vk\a}$ describing scattering directly between the leads and the bulk states. The rates appearing in this collision integral are significant only for $y$ values sufficiently close to the leads~\cite{SeeSM}.

\begin{figure}
  \centering
  \includegraphics[width=8.6cm]{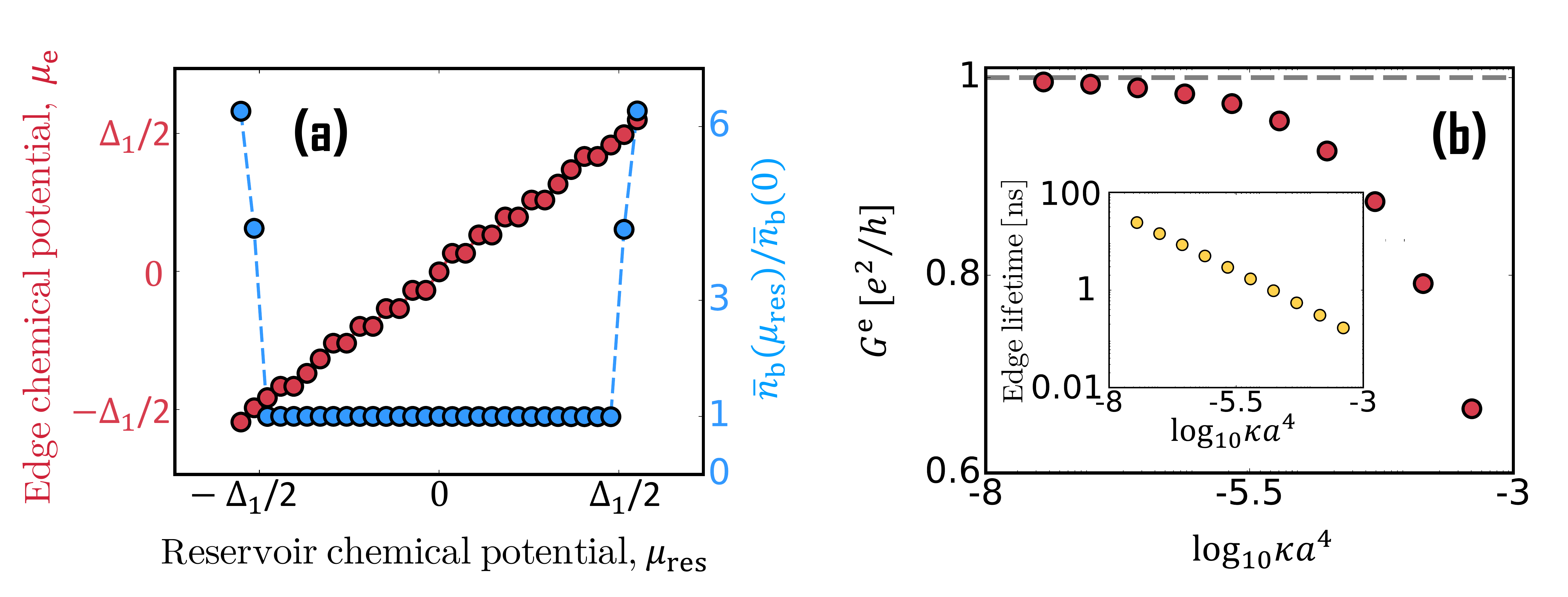}
  \caption{(Color online) (a) The effective chemical potential of the right edge, $\m_{\rm e}$, and the total normalized excitation density in the bulk, $\bar{n}_\yb(\m_{\rm res})/\bar{n}_\yb(\m_{\rm res}=0)$, for a system coupled to filtered Fermi reservoirs near the left and the right edges as a function of the common chemical potential of the two leads, $\mu_{\rm res}$. The system-lead couplings are taken to be $\overline\cJ_{\rm R},\overline\cJ_{\rm L}\eqa 2.4\Lm^{\rm e\to b}n_\yb$, and for their filter bandwidths we take $w = \hbar\Omega$. The chemical potential $\m_\ye$ is computed by fitting  the edge distribution to a quasi Fermi-Dirac distribution $f_{\rm QFD}$.
(b) The edge contribution to two-terminal conductance as a function of $\ka a^4$, for sample size $L_y=5$~$\rm \m m$, and Fermi velocity, $v_{\rm e}\eqa 10^{5}\frac{\rm m}{\rm sec}$. The conductance approaches the quantum limit, $e^2/h$, as $\ka a^4\to0$. Inset: The lifetime of the edge states due to edge-to-bulk scattering processes ($\ta_\ye$).  \label{fig:Transport}}
\end{figure}

The coupling strength between the reservoir and the edge states is characterized by $\overline{\cJ}_{\be}=\frac{1}{2k_R}\int_{-k_R}^{k_R}dk_x\cJ_{k_x;\be}^0$. When $\overline{\cJ}_{\be}\gg\Lm^{\rm e\to b}n_\yb$ (such that tunneling between the reservoir and the edge states dominates over scattering from the edge states to the bulk), we expect the distribution of the edge states to be described by the quasi Fermi-Dirac distribution $f_{\rm QFD}$, with an effective chemical potential $\mu_{\rm e}$ which is pinned to $\mu_{\rm res}$ \footnote[16]{The coupling to an energy filtered Fermi reservoir also affects the effective temperature and the $\dl$-parameter of the steady state.}. In contrast, we expect the total density of bulk excitations $\bar{n}_\yb=n_+ + n_-$ to remain constant when $\mu_{\rm res}$ is changed, as long as $\mu_{\rm res}$ remains within the Floquet gap. (The densities $n_+$ and $n_-$ correspond to the densities of electrons and holes in the upper and lower Floquet bands, respectively).
 In Fig.~\ref{fig:Transport}a we plot $\mu_{\rm e}$, as well as $\bar{n}_\yb(\m_{\rm res})/\bar n_\yb(\m_{\rm res}=0)$, as a function of $\m_{\rm res}$. The numerical results plotted in Fig.~\ref{fig:Transport}a indeed show the ``incompressible'' behavior of the bulk excitation density, and the pinning of $\mu_\ye$ on the edge to the chemical potential of the reservoir.

\paragraph{Transport signatures ---}
We consider a two-terminal transport measurement using a bar geometry, when a voltage bias $\D\m=\m^\yR_{\rm res}-\mu^\yL_{\rm res}$ is applied between the leads (see Fig.~\ref{fig:Geometry}b). The current through an FTI should in general have both bulk and edge contributions, characterized by a total conductance of the form $G=G^{\ye}+(L_x/L_y)\s^{\yb}_{yy}$ \footnote[15]{This formula applies also when $\s^\yb_{xy}\ne 0$ \cite{Moelter1998}}.
To estimate $G^{\ye}$, we consider an excess charge density on the right-moving edge due to occupation of edge modes with $\ve>0$.  We denote this quantity by $\D n_{\ye}$. The continuity equation for $\D n_{\ye}$ is given by $\dpa_t\D n_{\ye}= -v_\ye \dpa_y \D n_{\ye} - \bR{\D n_{\ye}-n_\ye}/\tau_{\ye}$, where $v_\ye$ is the edge velocity, $\tau_{\ye}$ is lifetime of the edge excitations, and $n_\ye$ is the density of excitations on the right-moving edge, far away from the leads, see Eq.~\eqref{eq:ModelResults2}. We define $\D n_\ye$ for the left movers accordingly. The lifetime $\tau_{\ye}$ is determined predominantly by edge-to-bulk scattering processes, such that $\ta_\ye\approx(\Lm^{\rm e\to b}n_\yb)\inv\sim\ka^{-\half}$. Assuming that the leads set the boundary conditions for $\D n_\ye$ at $y=0$ and $y=L_y$, for the right and left movers, correspondingly, we estimate the edge contribution to the two-terminal conductance: $G^{\ye}=(e^2/h)(1-\dl)e^{-L_y/{\ta_\ye v_\ye}}$~\cite{SeeSM}.
Fig.~\ref{fig:Transport}b displays the numerically obtained values of $\tau_{\ye}$, and the corresponding estimate for $G^{\ye}$ as a function of $\ka$. As $\ka \to0$, $\ta_{\rm e}$ increases and $\dl$ decreases; thus the conductance $G^\ye$ approaches the quantum limit $e^2/h$.

\paragraph{Discussion ---}
To estimate physically accessible values of $\ka$, we associate the phonon and photon mediated transitions with the typically observed hot electron lifetime, $\ta_{\rm he}\sim 0.1$~ps \cite{Schmuttenmaer1996,*Tanaka2003,*Wanga2013,*Niesner2014}, and the radiative recombination lifetime, $\ta_{\rm rr}\sim 0.1$~ns, respectively. For $A_\cR a^2\sim10^{-2}$, we then estimate $\ka_* a^4 \approx \frac{A_\cR a^2}{(2\p)^2}\frac{\ta_{\rm he}}{\ta_{\rm rr}}\sim 10^{-6}$. As seen in Fig.~\ref{fig:Transport}b, for this value of $\ka$ and a sample of the size $L_y= 1$~$\rm \m m$, $G^\ye$ is within a few percent of the quantized value.

The bulk contribution to the conductivity, $\s^\yb_{yy}$, will naturally depend on the material used to implement the FTI. Prominent candidates are CdTe/HgTe and InAs/GaSb heterostructures \cite{Lindner2011}, and honeycomb lattice materials such as transition-metal dichalcogenides \cite{Claassen2016}, and graphene \cite{Jotzu2014}. The low-temperature mobilities of these materials vary over a range of a few orders of magnitude \cite{Nafidi2013,*Safa2013,*Schmidt2015,*Xu2016}. Lower mobility samples, in which the bulk conductance is suppressed,  may be advantageous for measurements of $G^\ye$. We evaluate the bulk conductivity as $\s^\yb_{yy}\approx 2e \m n_\yb\approx (e^2/h)(\mu/\mu_\ast)(\ka/\kappa_\ast)^\half$, where $\mu$ is the mobility and $\mu_\ast=\frac{e}{2h\kappa^{1/2}_\ast}\sim 400$ $\rm\frac{cm^2}{V\cdot sec}$ \footnote[17]{The mobility includes both phonon and impurity scattering, see Supplementary Material.}
The bulk may also exhibit an anomalous Hall effect due to the non-zero Berry curvature of the Floquet bands. The Hall conductivity for low $\ka$ is of the order of $e^2/h$ and may be further renormalized by disorder \cite{Nagaosa2010}.

Our results demonstrate that the topological properties of the band structures of FTIs, and in particular the existence of edge states, can be manifested in an experimentally accessible transport measurement. To fully explore the possibilities offered by FTIs, other methods for detecting the edge states need to be developed. These  may include position dependent spectroscopic and magnetic probes~\cite{Dahlhaus2015, Nowack2013,Spanton2014,Yin2016}, as well as interference measurements between edge modes \cite{Ji2003}. Investigating the role of interparticle collisions in the driven system~\cite{Tsuji2008,Bilitewski2015a,Genske2015,Bukov2016} is also an important direction for future study.



\begin{acknowledgments}
\paragraph{Acknowledgements ---}
We thank Vladimir Kalnizky, Gali Matsman and Ari Turner for illuminating discussions, and David Cohen for technical support.  N. L. acknowledges support from  the European Research Council (ERC) under the European Union
Horizon 2020 Research and Innovation Programme (Grant Agreement No.
639172), from the People Programme
(Marie Curie Actions) of the European Union\textquoteright s Seventh
Framework Programme (FP7/2007\textendash 2013), under  REA Grant
Agreement No. 631696,  and from the Israeli Center of Research Excellence
(I-CORE) ``Circle of Light''. M. R. gratefully acknowledges the support of the European Research Council (ERC) under the European Union Horizon 2020 Research and Innovation Programme (Grant Agreement No. 678862), and the Villum Foundation. G.R. acknowledges support from the U. S. Army Research Office under grant number W911NF-16-1-0361, and from the IQIM, an NSF frontier center funded in part by the Betty and Gordon Moore Foundation. We also thank the Aspen Center for Physics, which is supported by National Science Foundation grant PHY-1607761 where part of the work was done.
\end{acknowledgments}

\clearpage


\section*{Supplementary material}

# Steady states and edge state transport in topological Floquet-Bloch systems − Supplementary material

Iliya Esin,[1] Mark S. Rudner,[2] Gil Refael,[3] and Netanel H. Lindner[1]

[1]*Physics Department, Technion, 320003 Haifa, Israel*

[2]*Center for Quantum Devices and Niels Bohr International Academy,*
*Niels Bohr Institute, University of Copenhagen, 2100 Copenhagen, Denmark*

[3]*Institute for Quantum Information and Matter, Caltech, Pasadena, CA 91125, USA*

(Dated: October 25, 2017)

## S1. DERIVATION OF THE BOLTZMANN KINETIC EQUATION

In this section we derive the Boltzmann equation for the system described in Eqs. (1)-(4) in the main text. We define the phase space distribution in the bulk, $f^{\rm b}_{\boldsymbol{k}\alpha}(\boldsymbol{r};t)$, [Eq. (8) in the main text] through Keldysh component of the Green function, $\mathcal{G}^K_\alpha = (1 - 2f^{\rm b}_\alpha)(\mathcal{G}^R_\alpha - \mathcal{G}^A_\alpha)$. As in the main text, $\alpha = +, -$ denotes the upper and lower Floquet bands, respectively. The Green's function $\mathcal{G}_\alpha$ satisfies the Dyson equation [S1, S2],

$$\left(G^{-1}_\alpha - \Sigma_\alpha\right) \circ \mathcal{G}_\alpha = \mathbb{1}. \tag{S1}$$

We write Eq. (S1) for Wigner-transformed functions, $\mathcal{G}_\alpha(\boldsymbol{k}, \boldsymbol{r}; \omega, t)$, $G_\alpha(\boldsymbol{k}, \boldsymbol{r}; \omega, t)$, and $\Sigma_\alpha(\boldsymbol{k}, \boldsymbol{r}; \omega, t)$. In this representation, $\circ \equiv \exp\left\{\frac{i}{2}\left(\overleftarrow{\partial}_{\boldsymbol{r}} \overrightarrow{\partial}_{\boldsymbol{k}} - \overleftarrow{\partial}_{\boldsymbol{k}} \overrightarrow{\partial}_{\boldsymbol{r}}\right) - \frac{i}{2}\left(\overleftarrow{\partial}_t \overrightarrow{\partial}_\omega - \overleftarrow{\partial}_\omega \overrightarrow{\partial}_t\right)\right\}$ denotes the Moyal operator (the arrows denote whether the derivative acts to the left or the right). The two-point functions over Keldysh time contour, $\mathcal{G}_\alpha$, $G_\alpha$, and $\Sigma_\alpha$ are arranged in a matrix form, for instance $\mathcal{G}_\alpha = \begin{pmatrix} \mathcal{G}^R_\alpha & \mathcal{G}^K_\alpha \\ 0 & \mathcal{G}^A_\alpha \end{pmatrix}$ [S3], and similarly for $G_\alpha$ and $\Sigma_\alpha$. Here $G_\alpha$ is the free propagator whose inverse is given by $[G^{-1}_\alpha]^{R/A} = \omega - \varepsilon_\alpha(\boldsymbol{k})/\hbar \pm i0^+$, $[G^{-1}_\alpha]^K \approx 0$ (note that $G_\alpha$ is independent of $\boldsymbol{r}$ and $t$). The full form of the self energy $\Sigma_\alpha$ appearing in Eq. (S1) will be given below.

The distribution of electrons in the edge states, $f^e_{k_x\beta}(t)$ (see main text below Eq. (8)), is defined through Keldysh component of the edge Green function $\tilde{\mathcal{G}}^K_\beta = (1 - 2f^e_\beta)(\tilde{\mathcal{G}}^R_\beta - \tilde{\mathcal{G}}^A_\beta)$. As in the main text, $\beta = {\rm R, L}$ denotes the right and left edge states. The edge Green's function $\tilde{\mathcal{G}}_\beta$ satisfies the Dyson equation [S1, S2],

$$\left(\tilde{G}^{-1}_\beta - \tilde{\Sigma}_\beta\right) \diamond \tilde{\mathcal{G}}_\beta = \mathbb{1}. \tag{S2}$$

where due to translation invariance in the $x$ direction, $\diamond \equiv \exp\left[-\frac{i}{2}\left(\overleftarrow{\partial}_t \overrightarrow{\partial}_\omega - \overleftarrow{\partial}_\omega \overrightarrow{\partial}_t\right)\right]$. The functions $\tilde{\mathcal{G}}_\beta(k_x; \omega, t)$, $\tilde{G}_\beta(k_x; \omega, t)$, and $\tilde{\Sigma}_\beta(k_x; \omega, t)$ are arranged in a matrix form in a similar manner to the bulk Green's functions. The free propagator for the edge states is given by $[\tilde{G}^{-1}_\beta]^{R/A} = \omega - \varepsilon_\beta(k_x)/\hbar \pm i0^+$ and $[\tilde{G}^{-1}_\beta]^K \approx 0$.

The bulk self energy $\Sigma_\alpha$ in Eq. (S1) has contributions from both bulk-bulk and bulk-edge scattering, $\Sigma_\alpha = \Sigma^{\rm bb}_\alpha + \Sigma^{\rm be}_\alpha$. We expand the self energy to the second order in the electron-boson coupling. At this order, the diagrams that lead to the self-energy appear in Fig. S1a. In the following derivation we will use the notation $|\phi^n_{\boldsymbol{k}\alpha}\rangle$ and $|\chi^n_{k_x\beta}(y)\rangle$ for the bulk and edge states appearing in the Fourier decomposition of Floquet states, see the discussion below Eq. (2) in the main text. For bulk-bulk scattering, the Wigner transform of the self energy $\Sigma^{\rm bb}_\alpha$ is given by

$$\begin{aligned}
\Sigma^{\rm bb}_{\alpha,\lambda}(\boldsymbol{k}, y; \omega, t) =& \frac{2\pi i}{\hbar} a^3 \int \frac{d\omega' d^3\boldsymbol{q}}{(2\pi)^4} \frac{1}{L_x L_y} \sum_{n,\alpha'\boldsymbol{k}'} \left|\sum_m \langle\phi^m_{\boldsymbol{k}\alpha}|\eta_\lambda|\phi^{m-n}_{\boldsymbol{k}'\alpha'}\rangle\right|^2 (2\pi)^2 \delta^{(2)}(\boldsymbol{k} - \boldsymbol{k}' - \boldsymbol{q}) \times \\
& \times \sum_{v,v' \in \{{\rm q,cl}\}} \gamma^v \mathcal{G}_{\alpha'}(\boldsymbol{k}', y; \omega - \omega' - n\Omega, t) \gamma^{v'} \left[D_{vv'}(\boldsymbol{q}, \omega') + D^*_{v'v}(-\boldsymbol{q}, -\omega')\right].
\end{aligned} \tag{S3}$$

Here $\gamma^{\rm q} = \frac{1}{2}\mathbb{1}$, $\gamma^{\rm cl} = \frac{1}{2}\sigma^x$ are the vertex fermionic matrices [S2]; the functions $D_{v,v'}$ are given by $D_{\rm cl,q} = \frac{1}{\omega - v_\lambda|\boldsymbol{q}| \pm i0^+}$, $D_{\rm q,cl} = D^*_{\rm cl,q}$, and $D_{\rm cl,cl} = \coth\left(\frac{\hbar\omega}{2k_BT}\right)(D_{\rm cl,q} - D_{\rm q,cl})$, is the free bosonic propagator; $\delta^{(2)}(\boldsymbol{k} - \boldsymbol{k}' - \boldsymbol{q})$ enforces momentum conservation in the two dimensional plane of the system. Expanding the self energy to leading order, we approximate $\mathcal{G}^{R/A}_\alpha \approx G^{R/A}_\alpha$.

For bulk-edge scattering, the Wigner transform of the self energy $\Sigma_\alpha^{\mathrm{be}}$ is given by

$$\Sigma_{\alpha,\lambda}^{\mathrm{be}}(\boldsymbol{k}, y; \omega, t) = \frac{2\pi i}{\hbar} a^3 \int \frac{d\omega' d^3\boldsymbol{q}}{(2\pi)^4} \frac{1}{L_x} \sum_{n,\beta' k'_x} \zeta_{n,\boldsymbol{k}\alpha}^{k'_x \beta'}(q_y, y)(2\pi)\delta^{(1)}(k_x - k'_x - q_x) \times$$
$$\times \sum_{v,v' \in \{q,cl\}} \gamma^v \tilde{\mathcal{G}}_{\beta'}(k'_x; \omega - \omega' - n\Omega, t)\gamma^{v'} \left[ D_{vv'}(\boldsymbol{q}, \omega') + D_{v'v}^*(-\boldsymbol{q}, -\omega') \right]. \tag{S4}$$

where we have defined

$$\zeta_{n,\boldsymbol{k}\alpha}^{k'_x \beta'}(q_y, y) = \int \frac{d\bar{k}_y}{a\pi} e^{2i\bar{k}_y y} \int d\bar{y} e^{-i(k_y + \bar{k}_y - q_y)\bar{y}} \sum_m \langle \phi_{\boldsymbol{k}+\bar{\boldsymbol{k}}_y \hat{\boldsymbol{y}}\alpha}^m | \eta_\lambda | \chi_{k'_x \beta'}^{m-n}(\bar{y}) \rangle \int d\bar{y}' e^{i(k_y - \bar{k}_y - q_y)\bar{y}'} \sum_m \langle \chi_{k'_x \beta'}^{m-n}(\bar{y}') | \eta_\lambda^\dagger | \phi_{\boldsymbol{k}-\bar{\boldsymbol{k}}_y \hat{\boldsymbol{y}}\alpha}^m \rangle. \tag{S5}$$

Since the the edge states are localized in $y$, $\zeta_{n,\boldsymbol{k}\alpha}^{k'_x \beta'}(y)$ has a compact support near the position of the edges.

The edge self energy $\tilde{\Sigma}_\beta$ in Eq. (S2) has contributions from both edge-bulk and edge-edge scattering, $\tilde{\Sigma}_\beta = \Sigma_\beta^{\mathrm{eb}} + \Sigma_\beta^{\mathrm{ee}}$. The Wigner transforms of these self energies read

$$\Sigma_{\beta,\lambda}^{\mathrm{eb}}(k_x; \omega, t) = \frac{2\pi i}{\hbar} a^3 \int \frac{d\omega' d^3\boldsymbol{q}}{(2\pi)^4} \frac{a}{L_y} \sum_{n,\alpha' \boldsymbol{k}' y} \tilde{\zeta}_{n,k_x \beta}^{\boldsymbol{k}'\alpha'}(q_y, y)(2\pi)\delta^{(1)}(k_x - k'_x - q_x) \times$$
$$\times \sum_{v,v' \in \{q,cl\}} \gamma^v \mathcal{G}_{\alpha'}(\boldsymbol{k}', y; \omega - \omega' - n\Omega, t)\gamma^{v'} \left[ D_{vv'}(\boldsymbol{q}, \omega') + D_{v'v}^*(-\boldsymbol{q}, -\omega') \right] \tag{S6}$$

$$\Sigma_{\beta,\lambda}^{\mathrm{ee}}(k_x; \omega, t) = \frac{2\pi i}{\hbar} a^3 \int \frac{d\omega' d^3\boldsymbol{q}}{(2\pi)^4} \frac{1}{L_x} \sum_{n,\beta' k'_x} \left| \sum_m \langle \chi_{k_x \beta}^m | \eta_\lambda | \chi_{k'_x \beta'}^{m-n} \rangle \right|^2 (2\pi)\delta^{(1)}(k_x - k'_x - q_x) \times$$
$$\times \sum_{v,v' \in \{q,cl\}} \gamma^v \tilde{\mathcal{G}}_{\beta'}(k'_x; \omega - \omega' - n\Omega, t)\gamma^{v'} \left[ D_{vv'}(\boldsymbol{q}, \omega') + D_{v'v}^*(-\boldsymbol{q}, -\omega') \right]. \tag{S7}$$

where

$$\tilde{\zeta}_{n,k_x \beta}^{\boldsymbol{k}'\alpha'}(q_y, y) = \int \frac{d\bar{k}_y}{a\pi} e^{2i\bar{k}_y y} \int d\bar{y} e^{i(k_y + \bar{k}_y + q_y)\bar{y}} \sum_m \langle \chi_{k'_x \beta'}^m(\bar{y}) | \eta_\lambda | \phi_{\boldsymbol{k}+\bar{\boldsymbol{k}}_y \hat{\boldsymbol{y}}\alpha}^{m-n} \rangle \int d\bar{y}' e^{-i(k_y - \bar{k}_y + q_y)\bar{y}'} \sum_m \langle \phi_{\boldsymbol{k}-\bar{\boldsymbol{k}}_y \hat{\boldsymbol{y}}\alpha}^{m-n} | \eta_\lambda^\dagger | \chi_{k'_x \beta'}^m(\bar{y}') \rangle. \tag{S8}$$

Eqs. (S1) and (S2) are the kinetic equations for the Floquet bulk and edge states, written in the matrix form. Writing the explicit expressions for the Keldysh matrices $G_\alpha, \mathcal{G}_\alpha$, and $\Sigma_\alpha$ in terms of their components, we arrive at three independent equations, that for the bulk read

$$\left([G_\alpha^{-1}]^R - \Sigma_\alpha^R\right) \circ \mathcal{G}_\alpha^R = \mathbb{1} \tag{S9a}$$

$$\left([G_\alpha^{-1}]^A - \Sigma_\alpha^A\right) \circ \mathcal{G}_\alpha^A = \mathbb{1} \tag{S9b}$$

$$\left([G_\alpha^{-1}]^R - \Sigma_\alpha^R\right) \circ \mathcal{G}_\alpha^K + \left([G_\alpha^{-1}]^K - \Sigma_\alpha^K\right) \circ \mathcal{G}_\alpha^A = 0. \tag{S9c}$$

Playing around with these equations, and neglecting derivatives of $\Sigma$, we arrive at

$$[G_\alpha^{-1}]^R \circ f_\alpha^{\mathrm{b}} - f_\alpha^{\mathrm{b}} \circ [G_\alpha^{-1}]^A \approx \frac{1}{2} \left[ \Sigma_\alpha^K - \left(\Sigma_\alpha^R - \Sigma_\alpha^A\right)(1 - 2f_\alpha^{\mathrm{b}}) \right]. \tag{S10}$$

This equation can be further simplified to the form of the Boltzmann equation (Eq. (9) in the main text) evaluating $[G_\alpha^{-1}]^R \circ f_\alpha^{\mathrm{b}} - f_\alpha^{\mathrm{b}} \circ [G_\alpha^{-1}]^A \approx i\partial_t f_\alpha^{\mathrm{b}} + (i/\hbar)\partial_{\boldsymbol{k}}\varepsilon_\alpha(\boldsymbol{k})\partial_{\boldsymbol{r}} f_\alpha^{\mathrm{b}}$, where we neglected higher order derivatives of $f$. The equation for the edge has a similar form, $[\tilde{G}_\beta^{-1}]^R \circ f_\beta^{\mathrm{e}} - f_\beta^{\mathrm{e}} \circ [\tilde{G}_\beta^{-1}]^A \approx i\partial_t f_\beta^{\mathrm{e}}$. The collision integral is proportional to the right hand side of Eq. (S10), in particular,

$$\mathcal{I} = -\frac{i}{2} \left[ \Sigma^K - \left(\Sigma^R - \Sigma^A\right)(1 - 2f) \right] \Big|_{\hbar\omega = \varepsilon_\alpha(\boldsymbol{k})}. \tag{S11}$$

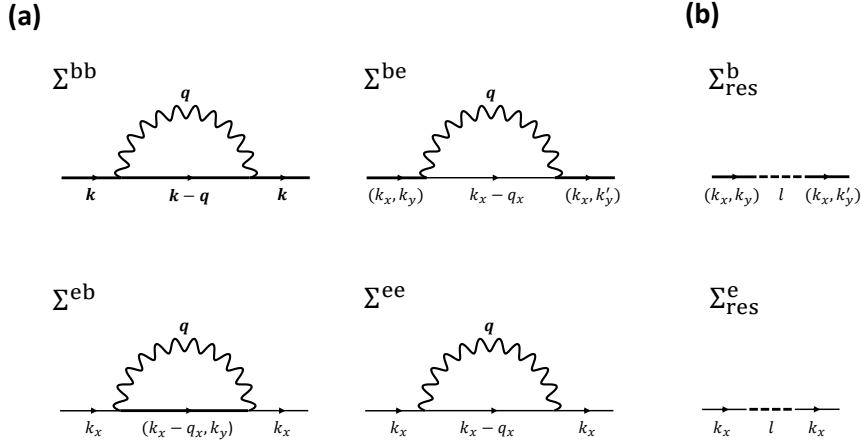

FIG. S1. Diagrams that lead to the self energy, (a) due to electron-boson collision processes; (b) due to system-lead tunneling. Bold and thin full lines correspond to bulk and edge propagators, respectively; wiggly and dashed lines denote the propagators of bosons and particles in the lead.

### S1.1.  Explicit expressions for the edge and the bulk collision integrals

Let us now present explicit expressions for the Boltzmann equation for the bulk and the edge states, for zero d.c. electric field. For completeness, we will assume reservoirs at finite temperature, $T$. Then both Boltzmann equations read

$$\partial_t f^{\mathrm{b}}_{\boldsymbol{k}\alpha} + v_{y,\alpha}(\boldsymbol{k})\partial_y f^{\mathrm{b}}_{\boldsymbol{k}\alpha} = \mathcal{I}^{\mathrm{bb}}_{\boldsymbol{k}\alpha}(y) + \mathcal{I}^{\mathrm{bR}}_{\boldsymbol{k}\alpha}(y) + \mathcal{I}^{\mathrm{bL}}_{\boldsymbol{k}\alpha}(y) \tag{S12a}$$

$$\partial_t f^{\mathrm{e}}_{k_x\beta} = \mathcal{I}^{\mathrm{ee}}_{k_x\beta} + \mathcal{I}^{\mathrm{eb}}_{k_x\beta}, \tag{S12b}$$

where $v_{y,\alpha}(\boldsymbol{k}) = \hbar^{-1}\partial_{k_y}\varepsilon_\alpha(\boldsymbol{k})$, and the collision integrals are

$$\mathcal{I}^{\mathrm{bb}}_{\boldsymbol{k}\alpha}(y) = \sum_{n,\boldsymbol{k}'\alpha'}\left[W^{\mathrm{bb},\boldsymbol{k}'\alpha'}_{n,\boldsymbol{k}\alpha}\mathscr{F}^{\mathrm{bb},\boldsymbol{k}'\alpha'}_{n,\boldsymbol{k}\alpha}(y) - W^{\mathrm{bb},\boldsymbol{k}\alpha}_{n,\boldsymbol{k}'\alpha'}\mathscr{F}^{\mathrm{bb},\boldsymbol{k}\alpha}_{n,\boldsymbol{k}'\alpha'}(y)\right] \tag{S13a}$$

$$\mathcal{I}^{\mathrm{be}}_{\boldsymbol{k}\alpha}(y) = \sum_{n,k'_x\beta'}\left[W^{\mathrm{be},k'_x\beta'}_{n,\boldsymbol{k}\alpha}(y)\mathscr{F}^{\mathrm{be},k'_x\beta'}_{n,\boldsymbol{k}\alpha}(y) - W^{\mathrm{eb},\boldsymbol{k}\alpha}_{n,k'_x\beta'}(y)\mathscr{F}^{\mathrm{eb},\boldsymbol{k}\alpha}_{n,k'_x\beta'}(y)\right] \tag{S13b}$$

$$\mathcal{I}^{\mathrm{ee}}_{k_x\beta} = \sum_{n,k'_x\beta'}\left[W^{\mathrm{ee},k'_x\beta'}_{n,k_x\beta}\mathscr{F}^{\mathrm{ee},k'_x\beta'}_{n,k_x\beta} - W^{\mathrm{ee},k_x\beta}_{n,k'_x\beta'}\mathscr{F}^{\mathrm{ee},k_x\beta}_{n,k'_x\beta'}\right] \tag{S13c}$$

$$\mathcal{I}^{\mathrm{eb}}_{k_x\beta} = \frac{a}{L_y}\sum_{n,\boldsymbol{k}'\alpha',y}\left[W^{\mathrm{eb},\boldsymbol{k}'\alpha'}_{n,k_x\beta}(y)\mathscr{F}^{\mathrm{eb},\boldsymbol{k}'\alpha'}_{n,k_x\beta}(y) - W^{\mathrm{be},k_x\beta}_{n,\boldsymbol{k}'\alpha'}(y)\mathscr{F}^{\mathrm{be},k_x\beta}_{n,\boldsymbol{k}'\alpha'}(y)\right]. \tag{S13d}$$

Where we defined, for brevity, $\mathscr{F}^{\varrho\varrho',\boldsymbol{k}'\alpha'}_{n,\boldsymbol{k}\alpha}(y) = [1 - f^{\varrho}_{\boldsymbol{k}\alpha}(y)]f^{\varrho'}_{\boldsymbol{k}'\alpha'}(y)N^{\boldsymbol{k}'\alpha'}_{n,\boldsymbol{k}\alpha} - f^{\varrho}_{\boldsymbol{k}\alpha}(y)[1 - f^{\varrho'}_{\boldsymbol{k}'\alpha'}(y)](1 + N^{\boldsymbol{k}'\alpha'}_{n,\boldsymbol{k}\alpha})$, $N^{\boldsymbol{k}'\alpha'}_{n,\boldsymbol{k}\alpha} = \left[\exp\left(\frac{\varepsilon_\alpha(\boldsymbol{k}) - \varepsilon_{\alpha'}(\boldsymbol{k}') + n\hbar\Omega}{k_{\mathrm{B}}T}\right) - 1\right]^{-1}$, and $W^{\varrho\varrho',\boldsymbol{k}'\alpha'}_{n,\boldsymbol{k}\alpha} = W^{\varrho\varrho',\boldsymbol{k}'\alpha'}_{\ell,n,\boldsymbol{k}\alpha} + W^{\varrho\varrho',\boldsymbol{k}'\alpha'}_{s,n,\boldsymbol{k}\alpha}$ for $\varrho$ and $\varrho' = \{\mathrm{b}, \mathrm{e}\}$ denoting the bulk and edge, respectively. The dependence on $y$ for the functions corresponding to the edge ($\varrho = \mathrm{e}$) should be omitted. Notice, that we have introduced additional indices for $W$ relative to the definition in the main text, where we showed the expression for bulk-to-bulk collisions only (i.e., $\varrho = \varrho' = \mathrm{b}$). Explicit expressions for these rates can be derived

from Eqs. (S3),(S4),(S6),(S7), yielding

$$W^{\mathrm{bb},\boldsymbol{k}'\alpha'}_{\lambda,n,\boldsymbol{k}\alpha} = \frac{2\pi}{\hbar}\frac{1}{L_x L_y}\int a^3 d^3 \boldsymbol{q} \left|\sum_m \langle \phi^m_{\boldsymbol{k}\alpha}|\eta_\lambda|\phi^{m-n}_{\boldsymbol{k}'\alpha'}\rangle\right|^2 \delta\left(\varepsilon_\alpha(\boldsymbol{k}) - \varepsilon_{\alpha'}(\boldsymbol{k}') - \hbar v_\lambda|\boldsymbol{q}| + n\hbar\Omega\right)\delta^{(2)}(\boldsymbol{k} - \boldsymbol{k}' + \boldsymbol{q}) \quad \text{(S14a)}$$

$$W^{\mathrm{be},k'_x\beta'}_{\lambda,n,\boldsymbol{k}\alpha}(y) = \frac{2\pi}{\hbar}\frac{1}{L_x}\int \frac{a^3 d^3 \boldsymbol{q}}{2\pi}\zeta^{k'_x\beta'}_{n,\boldsymbol{k}\alpha}(q_y,y)\delta\left(\varepsilon_\alpha(\boldsymbol{k}) - \varepsilon_{\beta'}(k'_x) - \hbar v_\lambda|\boldsymbol{q}| + n\hbar\Omega\right)\delta^{(1)}(k_x - k'_x + \boldsymbol{q}) \quad \text{(S14b)}$$

$$W^{\mathrm{ee},k'_x\beta'}_{\lambda,n,k_x\beta} = \frac{2\pi}{\hbar}\frac{1}{L_x}\int \frac{a^3 d^3 \boldsymbol{q}}{2\pi}\left|\sum_m \langle \chi^m_{k_x\beta}|\eta_\lambda|\chi^{m-n}_{k'_x\beta'}\rangle\right|^2 \delta\left(\varepsilon_\beta(k_x) - \varepsilon_{\beta'}(k'_x) - \hbar v_\lambda|\boldsymbol{q}| + n\hbar\Omega\right)\delta^{(1)}(k_x - k'_x + \boldsymbol{q}) \quad \text{(S14c)}$$

$$W^{\mathrm{eb},\boldsymbol{k}'\alpha'}_{\lambda,n,k_x\beta}(y) = \frac{2\pi}{\hbar}\frac{1}{L_x}\int \frac{a^3 d^3 \boldsymbol{q}}{2\pi}\tilde{\zeta}^{\boldsymbol{k}'\alpha'}_{n,k_x\beta}(q_y,y)\delta\left(\varepsilon_\beta(k_x) - \varepsilon_{\alpha'}(\boldsymbol{k}') - \hbar v_\lambda|\boldsymbol{q}| + n\hbar\Omega\right)\delta^{(1)}(k_x - k'_x + \boldsymbol{q}), \quad \text{(S14d)}$$

To obtain useful expressions for the simulation, we need to make assumptions on the profile of the edge wavefunctions, $|\chi^n_{k_x,\beta}\rangle$ as function of $y$. This will set closed expressions for $\zeta$-functions. We will assume the edge states are exponentially localized on the length on the order of the lattice spacing, for simplicity we take, $|\chi^n_{k_x,\beta}(y)\rangle = \delta_{y,0}|\bar{\chi}^n_{k_x,\beta}\rangle$. This approximation does not effect our numerical results, since in our simulation we discretize the $y$ direction on a length scale of the healing length which is much larger then the lattice scale (see Section S2). In this limit, we obtain

$$\zeta^{k'_x\beta'}_{n,\boldsymbol{k}\alpha}(q_y,y) = \delta_{y,0}\left|\sum_m \langle \phi^m_{\boldsymbol{k}\alpha}|\eta_\lambda|\bar{\chi}^{m-n}_{k'_x\beta'}\rangle\right|^2. \quad \text{(S15)}$$

Then the expressions for the rates factorize as

$$W^{\mathrm{bb},\boldsymbol{k}'\alpha'}_{\lambda,n,\boldsymbol{k}\alpha} = \frac{2\pi}{\hbar}\left|\sum_m \langle \phi^m_{\boldsymbol{k}\alpha}|\eta_\lambda|\phi^{m-n}_{\boldsymbol{k}'\alpha'}\rangle\right|^2 \times \rho^{\mathrm{b}}_\lambda\left(\varepsilon_\alpha(\boldsymbol{k}) - \varepsilon_{\alpha'}(\boldsymbol{k}') + n\hbar\Omega, \boldsymbol{k} - \boldsymbol{k}'\right) \quad \text{(S16a)}$$

$$W^{\mathrm{be},k'_x\beta'}_{\lambda,n,\boldsymbol{k}\alpha} = \frac{2\pi}{\hbar}\delta_{y,0}\left|\sum_m \langle \phi^m_{\boldsymbol{k}\alpha}|\eta_\lambda|\bar{\chi}^{m-n}_{k'_x\beta'}\rangle\right|^2 \times \rho^{\mathrm{e}}_\lambda\left(\varepsilon_\alpha(\boldsymbol{k}) - \varepsilon_{\beta'}(k'_x) + n\hbar\Omega, k_x - k'_x\right) \quad \text{(S16b)}$$

$$W^{\mathrm{ee},k'_x\beta'}_{\lambda,n,k_x\beta} = \frac{2\pi}{\hbar}\left|\sum_m \langle \chi^m_{k_x\beta}|\eta_\lambda|\chi^{m-n}_{k'_x\beta'}\rangle\right|^2 \times \rho^{\mathrm{e}}_\lambda\left(\varepsilon_\beta(k_x) - \varepsilon_{\beta'}(k'_x) + n\hbar\Omega, k_x - k'_x\right) \quad \text{(S16c)}$$

$$W^{\mathrm{eb},k'\alpha'}_{\lambda,n,k_x\beta} = \frac{2\pi}{\hbar}\delta_{y,0}\left|\sum_m \langle \bar{\chi}^m_{k_x\beta}|\eta_\lambda|\phi^{m-n}_{\boldsymbol{k}'\alpha'}\rangle\right|^2 \times \rho^{\mathrm{e}}_\lambda\left(\varepsilon_\alpha(\boldsymbol{k}) - \varepsilon_{\beta'}(k'_x) + n\hbar\Omega, k_x - k'_x\right), \quad \text{(S16d)}$$

where $\rho^{\mathrm{b}}_\lambda(\varepsilon,\boldsymbol{q}) = \frac{a^2}{L_x L_y}\frac{a\varepsilon\Theta(\varepsilon - \hbar v_\lambda|\boldsymbol{q}|)}{\pi\hbar v_\lambda\sqrt{\varepsilon^2 - \hbar^2 v_\lambda^2|\boldsymbol{q}|^2}}$ and $\rho^{\mathrm{e}}_\lambda(\varepsilon,q_x) = \frac{a}{L_x}\frac{a^2\varepsilon}{2\pi\hbar^2 v_\lambda^2}\Theta(\varepsilon - \hbar v_\lambda|q_x|)$; here $\Theta(\varepsilon)$ is the Heaviside step function.

### S1.2. Point coupling to a lead

We now derive the collision integral for the system-lead coupling (see Eq. (13) in the main text). We assume a large filtered lead with states, $|l\rangle$, and energies $\mathcal{E}_l$, characterized by a composite index, $l = \{k_x, \mathcal{E}_l, \nu\}$, were $\nu = \{\uparrow,\downarrow\}$, and $k_x \in [-\pi/a, \pi/a)$. We take the modes to provide a uniform density of states, independent of $k_x$ and $\nu$, throughout the filter window. We place the filer window centred around the energy $\hbar\Omega/2$ in the original conduction band, with the width of $\hbar\Omega$, i.e. $0 < \mathcal{E}_l < \hbar\Omega$. The right lead-reservoir coupling is assumed to have the form $J^{\mathrm{R}}_{lp} = J\delta_{yL_y}\delta_{\nu(p)\nu(l)}\delta_{k_x(p)k_x(l)}$. Here $p$ is a compact notation to label system states, $p = (k_x, y, \nu)$. The self energy due to system lead tunneling is represented diagrammatically in Fig. S1b. The Wigner transform of the self energy due to tunneling of a particle from a bulk state to the right lead, reads

$$\Sigma^{\mathrm{b}}_{\mathrm{res,R},\alpha}(\boldsymbol{k},y;\omega) = \frac{2\pi}{\hbar}\sum_{n,l}|J|^2\zeta^{n,\nu(l),\boldsymbol{k}\alpha}_{\mathrm{res,R}}(y)\delta_{k_x,k_x(l)}g_l(\omega + n\Omega), \quad \text{(S17)}$$

where

$$\zeta^{n,\nu,\boldsymbol{k}\alpha}_{\mathrm{res,R}}(y) = \frac{a}{\pi}\int dk'_y e^{2ik'_y(y - L_y)}\langle \phi^n_{\boldsymbol{k} + k'_y\hat{\boldsymbol{y}}\alpha}|\nu\rangle\langle\nu|\phi^n_{\boldsymbol{k} - k'_y\hat{\boldsymbol{y}}\alpha}\rangle. \quad \text{(S18)}$$

Here $g_l(\omega) = \begin{pmatrix} g_l^R(\omega) & g_l^K(\omega) \\ 0 & g_l^A(\omega) \end{pmatrix}$ is the lead two point Keldysh function, where $g_l^{R(A)}(\omega) = \frac{1}{\omega - \mathcal{E}_l/\hbar \pm i0^+}$, and $g_l^K(\omega) = \tanh\left(\frac{\hbar\omega - \mu_{\text{res}}}{2k_B T}\right)\left(g_l^R(\omega) - g_l^A(\omega)\right)$. In general, $\Sigma$ might have off-diagonal terms in the $\alpha$ and $\alpha'$ indices. Those terms vanish in the limit $1/(\Delta\varepsilon\tau_{\text{scat}}) \to 0$, where $\Delta\varepsilon$ is of the order of the Floquet gap and $\tau_{\text{scat}}$ is the average scattering time due to phonon scattering [S4]. The self energy of the right edge has a similar form,

$$\Sigma_{\text{res,R}}^{\text{ee}}(k_x;\omega) = \frac{2\pi}{\hbar}\sum_{n,l}|J|^2\left|\langle\chi_{k_x\text{R}}^n|L_y,\nu(l)\rangle\right|^2\delta_{k_x,k_x(l)}g_l(\omega + n\Omega). \tag{S19}$$

We employ Eq. (S11) to write the collision integral due to right reservoir tunneling,

$$\mathcal{I}_{\boldsymbol{k}\alpha}^{\text{b,res}} = \sum_n \mathcal{J}_{\boldsymbol{k}\alpha}^n\left[f_{\text{FD}}\left(\varepsilon_\alpha^n(\boldsymbol{k}) - \mu_{\text{res}}\right) - f_{\boldsymbol{k}\alpha}^{\text{b}}\right] \tag{S20a}$$

$$\mathcal{I}_{k_x\text{R}}^{\text{e,res}} = \sum_n \mathcal{J}_{k_x\text{R}}^n\left[f_{\text{FD}}\left(\varepsilon_\text{R}^n(k_x) - \mu_{\text{res}}\right) - f_{k_x\text{R}}^{\text{e}}\right], \tag{S20b}$$

where

$$\mathcal{J}_{\boldsymbol{k}\alpha}^n = \frac{2\pi}{\hbar}|J|^2\sum_\nu \zeta_{\text{res,R}}^{n,\nu(l),\boldsymbol{k}\alpha}(y)\mathcal{N}(\varepsilon_\alpha(\boldsymbol{k}) + n\hbar\Omega) \tag{S21a}$$

$$\mathcal{J}_{k_x\text{R}}^n = \frac{2\pi}{\hbar}|J|^2\sum_\nu\left|\langle\nu, L_y|\chi_{k_x\text{R}}^n\rangle\right|^2\mathcal{N}(\varepsilon_\text{R}(k_x) + n\hbar\Omega), \tag{S21b}$$

and $\mathcal{N}(\varepsilon)$ is a rectangular function around a single Floquet zone, describing the energy filtering window: $\mathcal{N}(\varepsilon) = 1$ if $0 < \varepsilon < \hbar\Omega$, $\mathcal{N}(\varepsilon) = 0$ otherwise.

### S1.3. Estimation of the effective recombination rates and scaling with the system size

Here we estimate the effective rates in Eq. (5) in the main text and analyze how they scale with the system size. We evaluate the effective rates from the rates in the microscopic model (Eqs. (S16)), employing the steady state distributions, $f_{\boldsymbol{k}\alpha}^{\text{b}}$, $f_{k_x\beta}^{\text{e}}$. We begin with the interband relaxation rate in the bulk, $\Lambda^{\text{inter}}$. We define the average relaxation rate for particles around minima of the Floquet band as $\overline{\mathcal{W}}^{\text{inter}} = \frac{\int d\boldsymbol{k}d\boldsymbol{k}' W_{s,\boldsymbol{k}+}^{\text{bb},\boldsymbol{k}'-}f_{\boldsymbol{k}'+}^{\text{b}}f_{\boldsymbol{k}+}^{\text{b}}}{\left(\int d\boldsymbol{k}f_{\boldsymbol{k}+}^{\text{b}}\right)^2}$. Then the rate to scatter to any of the hole states, per density of the holes, is given by $\Lambda^{\text{inter}} = L_xL_y\overline{\mathcal{W}}^{\text{inter}}$. At low excitation densities, $a^2 n_{\text{b}} \ll 1$, the population is significant near the band minima, at $\boldsymbol{k} = \boldsymbol{k}_R$, then we can approximate $\overline{\mathcal{W}}^{\text{inter}} \approx W_{s,\boldsymbol{k}_R+}^{\text{bb},\boldsymbol{k}_R-}$.

Next we find the recombination rate for particles in the upper Floquet band, $\mathcal{W}_{\boldsymbol{k}}^{\text{rec}}$. We observe that the momentum transfer in photon-mediated processes with typical energy $\hbar\Omega$, is of the order of $|\boldsymbol{q}| \sim \Omega/v_\ell$, which is rather small momentum compared to the scale over which the transition rates change. It follows that a state with momentum $\boldsymbol{k}$ can be scattered by a photon to a state $\boldsymbol{k}'$, within the radius $|\boldsymbol{q}|$. Since there are approximately $\sim \pi|\boldsymbol{q}|^2/\left(\frac{4\pi^2}{L_xL_y}\right) \approx \frac{L_xL_y}{4\pi}\frac{\Omega^2}{v_\ell^2}$ such states, we can approximate $\mathcal{W}_{\boldsymbol{k}}^{\text{rec}} \approx \frac{L_xL_y}{4\pi}\frac{\Omega^2}{v_\ell^2}W_{\ell,\boldsymbol{k}+}^{\text{bb},\boldsymbol{k}+}$.

We now employ the explicit expressions for the microscopic rates [Eq. (S16a)] to evaluate $\kappa = \frac{\Gamma^{\text{rec}}}{\Lambda^{\text{inter}}}$. Assuming the matrix elements of the electron-phonon and photon couplings in the active region are of the order of 1, we obtain $\overline{\mathcal{W}}^{\text{inter}} \approx g_s^2\frac{2a^3}{\hbar^2 v_s L_x L_y}$, and $\overline{\mathcal{W}}^{\text{rec}} \approx g_\ell^2\frac{a^3\Omega^2}{2\pi\hbar^2 v_\ell^3}$, where $\overline{\mathcal{W}}^{\text{rec}}$ is the average value of $\mathcal{W}_{\boldsymbol{k}}^{\text{rec}}$ within the resonance curve (see the inset in Fig. 1 in the main text). Now we employ, $\Gamma^{\text{rec}}\frac{A_\mathcal{R}}{(2\pi)^2}\overline{\mathcal{W}}_{\text{rec}}$, to approximate $\kappa$ as $\kappa \approx \frac{A_\mathcal{R}\Omega^2 v_s g_\ell^2}{8\pi^3 v_\ell^3 g_s^2}$.

To estimate the bulk-to-edge and edge-to-bulk processes let us define the average rates to scatter from the minima of the upper band to an edge state labeled by $k_x$ ($\overline{\mathcal{W}}_{k_x}^{\text{b}\to\text{e}}$), and from the edge to a maximum of the lower band ($\overline{\mathcal{W}}^{\text{e}\to\text{b}}$). The second process occurs essentially near $k_x = 0$, where most of the excitations accumulate. We define these rates as $\overline{\mathcal{W}}_{k_x}^{\text{b}\to\text{e}} = \frac{\int d\boldsymbol{k}W_{s,\boldsymbol{k}+}^{\text{be},k_x\text{R}}f_{\boldsymbol{k}+}^{\text{b}}}{\int d\boldsymbol{k}f_{\boldsymbol{k}+}^{\text{b}}}$, and $\overline{\mathcal{W}}^{\text{e}\to\text{b}} = \frac{\int d\boldsymbol{k}dk_x W_{s,k_x\text{R}}^{\text{eb},\boldsymbol{k}+}f_{\boldsymbol{k}+}^{\text{b}}f_{k_x\text{R}}^{\text{e}}}{\int d\boldsymbol{k}dk_x f_{\boldsymbol{k}+}^{\text{b}}f_{k_x\text{R}}^{\text{e}}}$ respectively. The parameter $\gamma^{\text{b}\to\text{e}}$ describes the scattering rate from the bulk to the edge per bulk density, per unit length. This process occurs essentially uniformly along the upper half of the edge, and can be approximated as $\gamma^{\text{b}\to\text{e}} \approx L_xL_y\int_0^{k_R}\frac{dk_x}{2\pi}\overline{\mathcal{W}}_{k_x}^{\text{b}\to\text{e}}$. Consequently, the edge-to-bulk rate per bulk density is given by $\Lambda^{\text{e}\to\text{b}} \approx L_xL_y\overline{\mathcal{W}}^{\text{e}\to\text{b}}$. Finally, the parameter $\gamma^{\text{e}\to\text{e}}$ can be estimated as $\gamma^{\text{e}\to\text{e}} \approx L_x\overline{\mathcal{W}}^{\text{e}\to\text{e}}$, where $\overline{\mathcal{W}}^{\text{e}\to\text{e}} = \frac{\int_0^{k_R}dk_xdk_x'W_{s,k_x\text{R}}^{\text{ee},k_x'\text{R}}f_{k_x\text{R}}^{\text{e}}f_{k_x'\text{R}}^{\text{e}}}{\left(\int_0^{k_R}dk_xf_{k_x\text{R}}^{\text{e}}\right)^2}$.

## S2. DETAILS OF THE NUMERICAL SIMULATIONS

Here we summarize the details of the numerical simulations. We discretize phase space using a grid of $N_{k_x} \times N_{k_y} = 50 \times 50$ sites in momentum space, and $N_y = 11$ sites in real space along the $y$-direction. For a physical system size of $L_y$, we define $\theta$ to be the sampling ratio of $k$-points, whereby $L_y = a\theta N_{k_y}$ (e.g. $\theta = 2$ means that we take every second $k$-point). Furthermore, the discretization step in the $y$-direction is given by $\Delta y = L_y/N_y = a\theta N_{k_y}/N_y$.

The discrete version of the steady state Boltzmann equation (see Eq. (9) in the main text) reads,

$$v_{y,\alpha}(\boldsymbol{k}) \frac{f_{\boldsymbol{k}\alpha}^{\mathrm{b}}(j+1) - f_{\boldsymbol{k}\alpha}^{\mathrm{b}}(j)}{a\theta N_{k_y}/N_y} = \mathcal{I}_{\boldsymbol{k}\alpha}^{\mathrm{tot}}(j), \tag{S22}$$

where $j$ is a discrete index indicating the position in the $y$ direction, i.e. $y = j\Delta y$. Here we denote by $\mathcal{I}_{\boldsymbol{k}\alpha}^{\mathrm{tot}}(j)$ the total contribution from the the bulk and the edge collision integrals. We now multiply Eq. (S22) by $\theta$ and $\Omega^{-1}$ to arrive at the dimensionless expression

$$\frac{v_{y,\alpha}(\boldsymbol{k})}{\Omega a} \frac{f_{\boldsymbol{k}\alpha}^{\mathrm{b}}(j+1) - f_{\boldsymbol{k}\alpha}^{\mathrm{b}}(j)}{N_{k_y}/N_y} = \frac{\theta \mathcal{I}_{\boldsymbol{k}\alpha}^{\mathrm{tot}}(j)}{\Omega}. \tag{S23}$$

We fix $\sqrt{\theta}g_s$ on the value that ensures that the step size $\Delta y = L_y/N_y$ is on the order of the healing length $\xi$, evaluated below (see also Fig. 3a in the main text).

Due to the discretization of real space in the $y$ direction, in the bulk-edge collision terms $\mathcal{I}_{\boldsymbol{k}\alpha}^{\mathrm{be}}(j)$ we replace $\delta_{\alpha,0}$ by $\frac{a}{\Delta y}\delta_{j,0}$ and $\delta_{\alpha,L_y}$ with $\frac{a}{\Delta y}\delta_{j,11}$, see Eqs. (S13b), (S16b), and (S16d). The prefactor, $a/\Delta y$, makes sure that the integrals of the new and the old $\delta$-functions are the same. Notice that Eq. (S23) is *not invariant* under the rescaling, $\theta \to \lambda\theta$, $g_s \to g_s/\sqrt{\lambda}$, $g_\ell \to g_\ell/\sqrt{\lambda}$ due to the $a/\Delta y$ prefactor in the bulk-to-edge collision integral. Therefore, both $L_y = a\theta N_{k_y}$ as well as $g_s$ and $\kappa = g_\ell/g_s$ have to specified when specifying the parameters of the simulation. We find the steady state solution to the Boltzmann equation (Eq. (9) in the main text) employing the Newton-Raphson method, for the parameters in Table S1.

| Simulation parameters | | | | | | | Physical parameters | |
|---|---|---|---|---|---|---|---|---|
| $A$ | $M$ | $B$ | $V_0$ | $\sqrt{\theta}g_s$ | $\theta$ | $N_y$ | $\Omega$ | $a$ |
| $0.2\hbar\Omega$ | $0.2\hbar\Omega$ | $-0.09\hbar\Omega$ | $0.2\hbar\Omega$ | $1\hbar\Omega$ | $2000$ | $11$ | $100$ THz | $5.6$ Å |
| $v_s$ | $v_\ell$ | $\omega_{\mathrm{D}}$ | $k_{\mathrm{B}}T$ | $J$ | $N_{k_x}$ | $N_{k_y}$ | $\tau_{\mathrm{he}}$ | $L_y$ |
| $0.0092a\Omega$ | $1a\Omega$ | $0.15\Omega$ | $0$ | $0.2\hbar\Omega$ | $50$ | $50$ | $0.1$ ps | $56$ μm |

TABLE S1. A list of the parameters used in the numerical simulations, and physical quantities corresponding to these simulation parameters.

Finally, we find the "hot electron" lifetime corresponding to the simulation parameters listed in Table S1. The coupling constant $g_s$ can be written in terms of the hot electron lifetime yielding $g_s \approx \hbar\sqrt{\frac{v_s}{2a\tau_{\mathrm{he}}}}$. Using Table S1 and taking $\Omega = 100$ THz, we obtain $\tau_{\mathrm{he}} \approx 0.1$ ps.

## S3. SPATIAL STRUCTURE OF THE STEADY STATE

Here we analyze the spatial structure of the bulk density in the steady state (see Fig. 3a in the main text) using a reaction-diffusion equation. We begin with the reaction-diffusion equation for the density of the bulk excitations in the steady state [S4],

$$D\partial_y^2 n_{\mathrm{b}}(y) = \Gamma^{\mathrm{rec}} - \Lambda^{\mathrm{inter}}n_{\mathrm{b}}^2(y). \tag{S24}$$

Here $D$ is the diffusion constant, given by $D \approx \bar{v}^2\tau$, where $\bar{v}$ is a typical velocity of the excited carriers and $\tau$ is the scattering time (dominated by phonon scattering). We solve this equation with boundary conditions at the edges of the system, such that the bulk current normal to the edges equals the rate of scattering into the edge states. For example, at the right edge of the system ($y = L_y$), the bulk current is given by $D\partial_y n_{\mathrm{b}}(L_y)$ and we set $D\partial_y n_{\mathrm{b}}(L_y) = -\gamma^{\mathrm{b}\to\mathrm{e}}n_{\mathrm{b}}(L_y) + \Lambda^{\mathrm{e}\to\mathrm{b}}n_{\mathrm{b}}(L_y)n_{\mathrm{e}}$. Assuming that the density of the bulk excitations is not strongly affected by the presence of the edge, we linearize Eq. (S24) around the bulk value, writing $\Delta n_{\mathrm{b}}(y) = n_{\mathrm{b}}(y) - n_{\mathrm{b}}^0$, where $n_{\mathrm{b}}^0 = n_{\mathrm{b}}(L_y/2)$. The diffusion equation then reads, $\partial_y^2\Delta n_{\mathrm{b}}(y) = \Delta n_{\mathrm{b}}(y)/\xi^2$ where $\xi = \sqrt{Dn_{\mathrm{b}}^0/(2\Gamma^{\mathrm{rec}})}$. Additionally,

we neglect the second term in the equation determining the boundary condition for the current, which is proportional to $\Lambda^{\text{e}\to\text{b}}$, since at low excitation density, the first term in this equation dominates (see the discussion a above Eq. (7) in the main text). Solving Eq. (S24) with the above boundary conditions we arrive at the expression for $\Delta\tilde{n}_{\text{b}}(y) = \frac{\Delta n_{\text{b}}(y)}{n_{\text{b}}^0}$,

$$\Delta\tilde{n}_{\text{b}}(y) = -\frac{\cosh((y - L_y/2)/\xi)}{\cosh(L_y/2\xi) + \frac{2\Gamma^{\text{rec}}\xi}{\gamma^{\text{b}\to\text{e}}n_{\text{b}}^0}\sinh(L_y/2\xi)}. \tag{S25}$$

Taking the limit $L_y \gg \xi \gg a$ we obtain

$$\Delta\tilde{n}_{\text{b}}(0) = -\frac{\gamma^{\text{b}\to\text{e}}n_{\text{b}}^0}{2\Gamma^{\text{rec}}\xi}. \tag{S26}$$

For our parameters (see Table S1) the estimate (S26) yields $\Delta\tilde{n}_{\text{b}}(0) \sim 10^{-3}$, which is in a good agreement with our results (see Fig. 3a in the main text).

The estimate in Eq. (S26) can be also obtained from Eq. (5a) in the main text. We first write $n_{\text{b}} = n_{\text{b}}^0 + \delta n_{\text{b}}$, where $n_{\text{b}}^0$ is the solution absent the bulk-edge scattering, and expand Eq. (5a) to first order in $\delta n_{\text{b}}$. We then assume that the excess number of particles excited due to edge-to-bulk scattering, $\delta N_{\text{b}} = \delta n_{\text{b}}L_x L_y$, accumulate only on a strip of width $\xi$ near each edge. The excess density in this strip is then given by Eq. (S26).

## S4. CONDUCTANCE IN TWO-PROBE SETUP

In this section we derive the conductance in the two-terminal setup (see Fig. 2b in the main text). The left and the right leads with chemical potentials $\mu_{\text{res}}^{\text{L}} = eV$ and $\mu_{\text{res}}^{\text{R}} = 0$ are connected at $y = 0$ and $y = L_y$ respectively. The total current in this geometry has two contributions from the bulk and the edge currents. The bulk contribution is computed from Ohm's law, with a conductivity tensor which we compute in Sec. 1. The conductivity tensor has both diagonal and off-diagonal components arising from the Berry curvature of the Floquet-Bloch bands. The edge current is derived in Sec. 2 from the continuity equation, taking into account scattering processes discussed in the main text.

### S4.1. Bulk contribution to the conductance

We first derive the conductivity tensor for the bulk of a FTI. The conductivity tensor has both longitudinal, $\sigma_{yy}^{\text{b}}$, and transversal, $\sigma_{xy}^{\text{b}}$, contributions. To find them in the clean limit, we write the Boltzmann equation for the bulk (see Eq. (9) in main text) in the relaxation time approximation. To find the response to external field, we include the coupling of the electric field to momentum gradients, $\partial_t f_{\boldsymbol{k}} \sim (e/\hbar)\boldsymbol{E} \cdot \partial_{\boldsymbol{k}} f_{\boldsymbol{k}}$. In the steady state, the Boltzmann equation for the electrons and the holes reads,

$$\boldsymbol{v}_+(\boldsymbol{k}) \cdot \partial_{\boldsymbol{r}} f_{\boldsymbol{k}+}^{\text{b}}(\boldsymbol{r}) - (e/\hbar)\boldsymbol{E} \cdot \partial_{\boldsymbol{k}} f_{\boldsymbol{k}+}^{\text{b}}(\boldsymbol{r}) = -\delta f_{\boldsymbol{k}+}^{\text{b}}(\boldsymbol{r})/\tau_{\boldsymbol{k}+} \tag{S27a}$$

$$\boldsymbol{v}_-(\boldsymbol{k}) \cdot \partial_{\boldsymbol{r}} \bar{f}_{\boldsymbol{k}-}^{\text{b}}(\boldsymbol{r}) - (e/\hbar)\boldsymbol{E} \cdot \partial_{\boldsymbol{k}} \bar{f}_{\boldsymbol{k}-}^{\text{b}}(\boldsymbol{r}) = -\delta \bar{f}_{\boldsymbol{k}-}^{\text{b}}(\boldsymbol{r})/\tau_{\boldsymbol{k}-}. \tag{S27b}$$

Here $\delta f_{\boldsymbol{k}+}^{\text{b}}(\boldsymbol{r}) = f_{\boldsymbol{k}+}^{\text{b}}(\boldsymbol{r}) - f_{\boldsymbol{k}+}^{\text{b},0}(\boldsymbol{r})$ is the deviation from the "local steady-state" distribution, $f_{\boldsymbol{k}\alpha}^{\text{b},0}$, satisfying $\mathcal{I}_{\boldsymbol{k}\alpha}^{\text{bb}}\left\{f_{\boldsymbol{k}\alpha}^{\text{b},0}\right\} = 0$. The velocity vector contains both the band velocity and the anomalous velocity components, namely $\boldsymbol{v}_\alpha(\boldsymbol{k}) = \hbar^{-1}\left[\partial_{\boldsymbol{k}}\varepsilon_\alpha(\boldsymbol{k}) - e\boldsymbol{E} \times \boldsymbol{\mathcal{F}}_{\alpha\boldsymbol{k}}\right]$, where $\boldsymbol{\mathcal{F}}_{\alpha\boldsymbol{k}}$ is the Berry curvature averaged over one period, given by $\boldsymbol{\mathcal{F}}_{\alpha\boldsymbol{k}} = -i\hat{\boldsymbol{z}}\text{Tr}\left\{P_{\alpha\boldsymbol{k}} \cdot \left[\partial_{k_x}P_{\alpha\boldsymbol{k}}, \partial_{k_y}P_{\alpha\boldsymbol{k}}\right]\right\}$. Here $P_{\alpha\boldsymbol{k}} = |\phi_{\boldsymbol{k}\alpha}^0\rangle\langle\phi_{\boldsymbol{k}\alpha}^0|$ is the projector to an eigenstate $|\phi_{\boldsymbol{k}\alpha}^0\rangle$. The relaxation time is approximated by $\tau_{\boldsymbol{k}\alpha}^{-1} = \frac{\delta\mathcal{I}_{\boldsymbol{k}\alpha}^{\text{bb}}}{\delta f_{\boldsymbol{k}\alpha}^{\text{b}}}\Big|_{f_{\boldsymbol{k}\alpha}^{\text{b},0}}$. Notice that in the clean case the relaxation time is determined by electron-phonon scattering. The solution to Eqs. (S27) to leading order in $\delta f^{\text{b}}$ and derivatives of $f^{\text{b}}$ reads

$$f_{\boldsymbol{k}+}^{\text{b}}(\boldsymbol{r}) \approx f_{\boldsymbol{k}+}^{\text{b},0}(\boldsymbol{r}) - \tau_{\boldsymbol{k}+}\boldsymbol{v}_+(\boldsymbol{k}) \cdot \partial_{\boldsymbol{r}} f_{\boldsymbol{k}+}^{\text{b}}(\boldsymbol{r}) + \tau_{\boldsymbol{k}+}(e/\hbar)\boldsymbol{E} \cdot \partial_{\boldsymbol{k}} f_{\boldsymbol{k}+}^{\text{b}}(\boldsymbol{r}) \tag{S28a}$$

$$\bar{f}_{\boldsymbol{k}-}^{\text{b}}(\boldsymbol{r}) \approx \bar{f}_{\boldsymbol{k}-}^{\text{b},0}(\boldsymbol{r}) - \tau_{\boldsymbol{k}-}\boldsymbol{v}_-(\boldsymbol{k}) \cdot \partial_{\boldsymbol{r}} \bar{f}_{\boldsymbol{k}-}^{\text{b}}(\boldsymbol{r}) + \tau_{\boldsymbol{k}-}(e/\hbar)\boldsymbol{E} \cdot \partial_{\boldsymbol{k}} \bar{f}_{\boldsymbol{k}-}^{\text{b}}(\boldsymbol{r}). \tag{S28b}$$

With this form for the electron and hole distributions, we can find the current density, given as

$$\boldsymbol{J} = \sum_{\alpha=\pm}\int \frac{d^2\boldsymbol{k}}{(2\pi)^2}e\boldsymbol{v}_\alpha(\boldsymbol{k})f_{\boldsymbol{k}\alpha}^{\text{b},0}. \tag{S29}$$

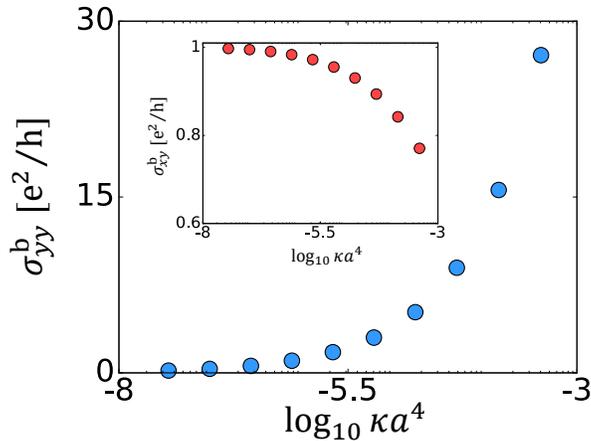

FIG. S2. The longitudinal ($\sigma_{yy}^{\mathrm{b}}$) and the transversal ($\sigma_{xy}^{\mathrm{b}}$) components of the bulk conductivity tensor vs $\kappa a^4$. The longitudinal component contains both the phonon contribution and the impurity contribution, for the typical value $\mu_{\mathrm{imp}} = 1000 \; \frac{\mathrm{cm}^2}{\mathrm{V \cdot sec}}$. The transversal component (inset) includes only the intrinsic contribution.

The conductivity tensor is defined by $\sigma_{ij}^{\mathrm{b}} = \frac{\partial J_i}{\partial E_j}$. Assuming an isotropic system, it is enough to find only two components, $\sigma_{yy}^{\mathrm{b}}$ and $\sigma_{xy}^{\mathrm{b}}$. The expression for the longitudinal component reads

$$\sigma_{\mathrm{ph},yy}^{\mathrm{b}} = 2\frac{e^2}{h} \int d^2\boldsymbol{k} \left( h^{-1}\partial_{k_y} \varepsilon_+(\boldsymbol{k}) \cdot \partial_{k_y} f_{\boldsymbol{k}_+}^{\mathrm{b},0} \right) \tau_{\boldsymbol{k}_+}. \tag{S30}$$

Here we assumed particle hole symmetry; the label "ph" denotes the contribution from phonon scattering. The transversal component reads

$$\sigma_{xy}^{\mathrm{b}} = \frac{e^2}{h} \frac{1}{2\pi} \sum_{\alpha=\pm} \int d^2\boldsymbol{k} (\hat{\boldsymbol{z}} \cdot \boldsymbol{\mathcal{F}}_{\boldsymbol{k}\alpha}) f_{\boldsymbol{k}\alpha}^{\mathrm{b},0}. \tag{S31}$$

In more realistic models, including disorder, the conductivity tensor may have additional contributions due to impurity scattering. We approximate the impurity contribution to longitudinal conductivity by $\sigma_{\mathrm{imp},yy}^{\mathrm{b}} = 2en_{\mathrm{b}}\mu_{\mathrm{imp}}$, where $\mu_{\mathrm{imp}}$ is the typical mobility due to impurity scattering. The total longitudinal conductivity from both contributions is found from Matthiessen's rule, $1/\sigma_{yy}^{\mathrm{b}} = 1/\sigma_{\mathrm{ph},yy}^{\mathrm{b}} + 1/\sigma_{\mathrm{imp},yy}^{\mathrm{b}}$. Fig. S2 shows the longitudinal and the transverse conductivities vs. $\kappa a^4$ for $\mu_{\mathrm{imp}} = 1000 \; \frac{\mathrm{cm}^2}{\mathrm{V \cdot sec}}$.

In principle, the scattering from impurities may affect the transverse component as well due to skew scattering and side jump processes [S5]. Here we do not consider these effects, focusing on the intrinsic contribution alone. A full analysis of skew scattering and side jumps in the Floquet system is an interesting subject for future work.

Now, we are ready to estimate the current in the two-probe geometry. The current in the bulk can be computed from the electric potential, $V(x, y)$, by Ohm's law, $J_i = \sigma_{ij}^{\mathrm{b}}\partial_i V$. To find $V(x, y)$, one needs to solve the Laplace equation, $\nabla^2 V = 0$, with appropriate boundary conditions. Along the open edges we require the currents to be zero in the normal to the edges direction, $\hat{\boldsymbol{n}}$, i.e. $\hat{n}_i \sigma_{ij}^{\mathrm{b}}\partial_i V = 0$. Near the leads we apply Dirichlet boundary conditions, i.e., $V(x, 0) = V$, and $V(x, L_y) = 0$. The current distribution in such a geometry is found in Ref [S6]. According to this result, the Hall contribution to the total current vanishes, leaving us with the expression for the conductance, $G^{\mathrm{b}} = (L_x/L_y)\sigma_{yy}^{\mathrm{b}}$.

### S4.2. Edge contribution to the conductance

To find the contribution to the conductance from the edge states, we write the continuity equation for each of the edges. The continuity equation is similar to Eq. (5b) in the main text, albeit with two changes: (i) the particle-hole symmetry is not valid anymore, leading to a different density of electrons ($\Delta n_{\mathrm{e}}$) and holes ($\Delta h_{\mathrm{e}}$) on the edge, when the chemical potentials are not at $\varepsilon = 0$. The densities of electrons and holes are evaluated as integrals over the edge distribution shifted by the chemical potential, $\mu_{\mathrm{e}}$, i.e. $\Delta n_{\mathrm{e}} = \int_0^{k_R} \frac{dk_y}{2\pi} f_{k_y}^{\mathrm{e}}(\mu_{\mathrm{e}})$, and $\Delta h_{\mathrm{e}} = \int_{-k_R}^0 \frac{dk_y}{2\pi} (1 - f_{k_y}^{\mathrm{e}}(\mu_{\mathrm{e}}))$.

The steady state distribution of the edge states can be fitted to a "quasi Fermi Dirac" distribution, $f_{k_y}^e(\mu_e) \approx f_{QFD}(\hbar v_e k_y - \mu_e) = (1-\delta)f_{FD}(\hbar v_e k_y - \mu_e) + \frac{1}{2}\delta$. Then to leading order in $\mu_e$ we obtain, $\Delta n_e \approx n_e + \frac{\mu_e}{4\pi\hbar v_e}$, and $\Delta h_e \approx n_e - \frac{\mu_e}{4\pi\hbar v_e}$, where $n_e$ is the steady state solution found in Eq. (7) in the main text. (ii) The density might not be uniform along the edge due to the difference in the chemical potentials near the leads, and due to edge-bulk scattering which can change $\Delta n_e$. To capture these features, we include a new gradient term. The continuity equation for the right edge in the steady state then reads

$$v_e \partial_y \Delta n_e = \gamma^{b\to e} n_b - \Lambda^{e\to b} n_b \Delta n_e - \gamma^{e\to e} \Delta n_e \Delta h_e, \tag{S32}$$

where we have approximated the edge velocity by a constant, $v_e$. We then expand Eq. (S32) to leading order in $\mu_e$, to obtain

$$v_e \partial_y \mu_e = -\mu_e/\tau_e + \mathcal{O}(\mu_e^2), \tag{S33}$$

where $\tau_e = (\Lambda^{e\to b} n_b)^{-1}$. We solve this equation with the boundary condition near the left lead $(y = 0)$, $\mu_e = eV$, to obtain $\mu_e(y) = eVe^{-y/v_e\tau_e}$. The edge contribution to the conductance then found from $G^e = \frac{\partial I}{\partial V}\big|_{y=L_y} = \frac{\partial}{\partial V}\int_{-k_R}^{k_R}\frac{dk_y}{2\pi}ev_e f_{QFD}(\hbar v_e k_y - \mu_e(L_y))$, yielding $G^e = (e^2/h)(1-\delta)e^{-L_y/v_e\tau_e}$.

To check this result, we have numerically evaluated the lifetime of the edge in the steady state (see inset in Fig. 4b in the main text). To this end we linearize Eq. (S12b) with a shifted chemical potential around the steady state at half filling (in the absence of leads), $f_{k_x R}^e = f_{k_x R}^{e,0} + \delta f_{k_x R}^e$, where $\delta f_{k_x R}^e = \frac{\partial f_{k_x R}^e}{\partial \mu_e}\big|_{\mu_e=0}\mu_e$. Then

$$\delta \dot{f}_{k_x R}^e = \sum_{k'\alpha'}\left[W_{k'\alpha'}^{be,k_x R} f_{k'\alpha'}^{b,0} \bar{f}_{k_x R}^e - W_{k_x R}^{eb,k'\alpha'} f_{k_x R}^e \bar{f}_{k'\alpha'}^{b,0}\right] + \sum_{k'_x}\left[W_{k'_x R}^{ee,k_x R} f_{k'_x R}^e \bar{f}_{k_x R}^e - W_{k_x R}^{ee,k'_x R} f_{k_x R}^e \bar{f}_{k'_x R}^e\right], \tag{S34}$$

where $\bar{f} = 1 - f$. Keeping only linear terms in delta f, we obtain

$$\delta \dot{f}_{k_x R}^e = -\sum_{k'\alpha'}\left[W_{k'\alpha'}^{be,k_x R} f_{k'\alpha'}^{b,0} + W_{k_x R}^{eb,k'\alpha'} \bar{f}_{k'\alpha'}^{b,0}\right]\delta f_{k_x R}^e - \sum_{k'_x}\left[W_{k'_x R}^{ee,k_x R} f_{k'_x R}^{e,0} + W_{k_x R}^{ee,k'_x R}\bar{f}_{k'_x R}^{e,0}\right]\delta f_{k_x R}^e +$$
$$+ \sum_{k'_x}\left[W_{k'_x R}^{ee,k_x R} f_{k_x R}^{e,0} + W_{k_x R}^{ee,k'_x R}\bar{f}_{k_x R}^{e,0}\right]\delta f_{k'_x R}^e + \mathcal{O}(\delta f^2). \tag{S35}$$

We define the lifetime, $\tau_e$, as $\sum_{k_x>0}\delta \dot{f}_{k_x,R}^e = -\frac{1}{\tau_e}\sum_{k_x>0}\delta f_{k_x R}^e$. We sum Eq. (S35) over the momentum $k_x > 0$. In the limit of small shift of the lead chemical potential, $\mu_e \to 0$, $\delta f_{k_x R} = \delta f_{-k_x R}$, due to particle-hole symmetry. Then the edge-to-edge contributions (the second and third terms in Eq. (S35)) cancel out and we remain with

$$\tau_e = \frac{\sum_{k_x>0}\delta f_{k_x R}^e}{\sum_{k_x>0}\sum_{k'\alpha'}\left[W_{k'\alpha'}^{be,k_x R} f_{k'\alpha'}^{b,0} + W_{k_x R}^{eb,k'\alpha'}\bar{f}_{k'\alpha'}^{b,0}\right]\delta f_{k_x R}^e}. \tag{S36}$$

The plot of $\tau_e$ is shown in the inset of Fig. 4 in the main text. It is in a good agreement to the phenomenological expression we obtained in the beginning of this section.

### S4.3. Dependence of the chemical potential on the lead-system coupling

In this section we study how the strength of the coupling to the lead ($J$ in Eq. (12) in the main text) affects the position of the chemical potential of the edge, $\mu_e$. Here we consider the part of the edge that is uniformly coupled to a lead (referring to Fig. 2b in the main text, we discuss the edges along the $x$ direction), hence the gradient term vanishes. We denote the shift of the lead chemical potential by $\mu_{res}$, and below assume $\mu_{res} > 0$. The continuity equations describing the electron and hole excitations at the right edge ($\Delta n_e$ and $\Delta h_e$, respectively), at $y = L_y$, are then

$$\Delta \dot{n}_e = \gamma^{b\to e} n_b - \Lambda^{e\to b} n_b \Delta n_e - \gamma^{e\to e}\Delta n_e \Delta h_e - \bar{\mathcal{J}}_R(\Delta n_e - n_{res}) \tag{S37a}$$

$$\Delta \dot{h}_e = \gamma^{b\to e} n_b - \Lambda^{e\to b} n_b \Delta h_e - \gamma^{e\to e}\Delta n_e \Delta h_e - \bar{\mathcal{J}}_R \Delta h_e. \tag{S37b}$$

Here $\bar{\mathcal{J}}_R$ is an average rate of edge-to-lead tunneling processes estimated as $\bar{\mathcal{J}}_R = \frac{1}{2k_R}\int_{-k_R}^{k_R}dk_x \mathcal{J}_{k_x R}^0$, and $n_{res} = \frac{\mu_{res}}{2\pi\hbar v_e}$ is the density of edge states integrated over the energy interval $0 < \varepsilon < \mu_{res}$, see Fig. S3. When $\mu_{res} = 0$, the system

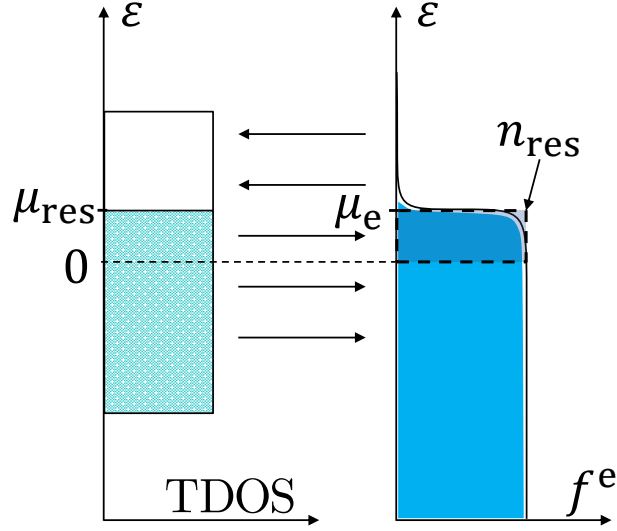

FIG. S3. Tunneling processes between the lead and the edge, when the chemical potential of the lead is shifted form the center of the band, at $\varepsilon = 0$. $n_{\text{res}}$ is the density of edge states integrated over the energy interval $0 < \varepsilon < \mu_{\text{res}}$.

is approximately particle-hole symmetric. The densities of the electrons ($\Delta n_{\text{e}}$) and holes ($\Delta h_{\text{e}}$) are then both equal to $n_{\text{e}}$, which is found in the main text (see Eq. (7)). When $\mu_{\text{res}}$ is slightly shifted from $\varepsilon = 0$, the densities of electrons and holes are shifted by $\delta n_{\text{e}} = \Delta n_{\text{e}} - n_{\text{e}}$ and $\delta h_{\text{e}} = \Delta h_{\text{e}} - n_{\text{e}}$, respectively. The steady state solution to Eqs. (S37a) and (S37b) (setting $\Delta \dot{n}_{\text{e}} = \Delta \dot{h}_{\text{e}} = 0$), to leading order in $\delta n_{\text{e}}$ and $\delta h_{\text{e}}$, then reads,

$$\delta n_{\text{e}} = \frac{\bar{\mathcal{J}}_{\text{R}} \left( \gamma^{\text{e} \to \text{e}} n_{\text{e}} + \Lambda^{\text{e} \to \text{b}} n_{\text{b}} + \bar{\mathcal{J}}_{\text{R}} \right)}{\left( \gamma^{\text{e} \to \text{e}} n_{\text{e}} + \Lambda^{\text{e} \to \text{b}} n_{\text{b}} + \bar{\mathcal{J}}_{\text{R}} \right)^2 - \left( \gamma^{\text{e} \to \text{e}} n_{\text{e}} \right)^2} n_{\text{res}} \tag{S38a}$$

$$\delta h_{\text{e}} = -\frac{\bar{\mathcal{J}}_{\text{R}} \gamma^{\text{e} \to \text{e}} n_{\text{e}}}{\left( \gamma^{\text{e} \to \text{e}} n_{\text{e}} + \Lambda^{\text{e} \to \text{b}} n_{\text{b}} + \bar{\mathcal{J}}_{\text{R}} \right)^2 - \left( \gamma^{\text{e} \to \text{e}} n_{\text{e}} \right)^2} n_{\text{res}}. \tag{S38b}$$

We estimate the position of the effective chemical potential of the edge by $\mu_{\text{e}} \approx 2\pi\hbar v_{\text{e}}(\delta n_{\text{e}} - \delta h_{\text{e}})$, then

$$\mu_{\text{e}} = \mu_{\text{res}} \frac{\bar{\mathcal{J}}_{\text{R}}}{\Lambda^{\text{e} \to \text{b}} n_{\text{b}} + \bar{\mathcal{J}}_{\text{R}}}. \tag{S39}$$

When $\bar{\mathcal{J}}_{\text{R}} \gg \Lambda^{\text{e} \to \text{b}}$ the chemical potential of the edge and the reservoir are equal, see Fig. 4a in the main text.

## S5. FITS OF THE STEADY STATES TO FERMI DIRAC AND QUASI FERMI DIRAC DISTRIBUTIONS

In this section we discuss the fit of the bulk distribution to a double Fermi function, and the fit of the edge to a quasi Fermi-Dirac distribution.

### S5.1. Distribution of the bulk

The steady state of the bulk is described by two separate Fermi-Dirac distributions, one for electrons in the upper Floquet band ($+$), and one for electrons in the lower Floquet band ($-$). The chemical potentials of these distributions, $\mu_+$ and $\mu_-$, are related by particle hole symmetry, $f^{\text{b}}_{\boldsymbol{k}+} = 1 - f^{\text{b}}_{\boldsymbol{k}-}$, which implies $\mu_+ = -\mu_- \equiv \mu_{\text{b}}$. Note that for any $\kappa > 0$, the chemical potentials are shifted away from $\varepsilon = 0$. Particle hole symmetry further implies that the effective temperatures, $T_{\text{b}}$, of the distributions in the both bands must be equal. Fig. S4a displays the least mean

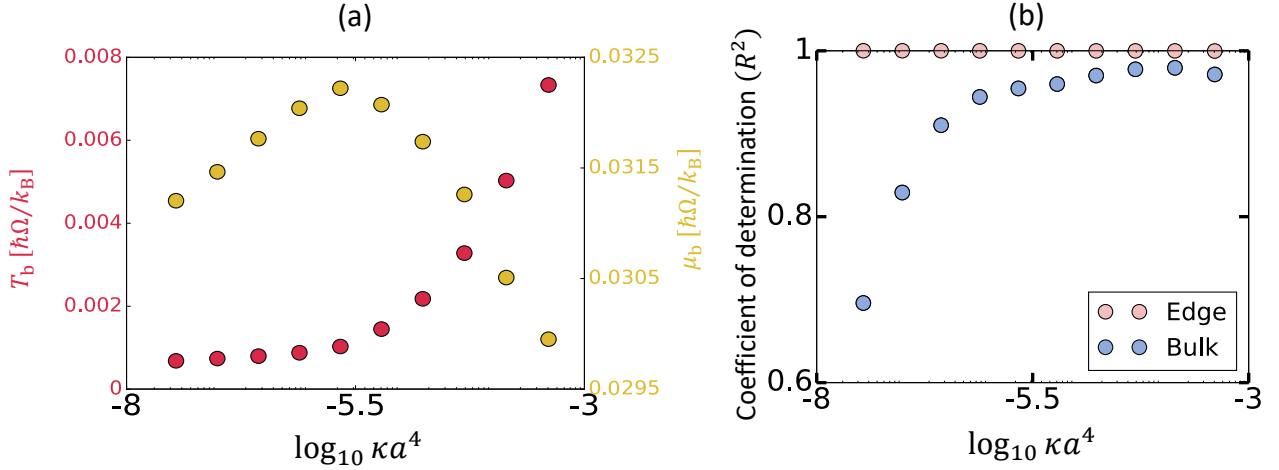

FIG. S4. (a) Effective parameters of the Fermi-Dirac distribution that fit the upper Floquet band steady state. (b) A coefficient of determination ($R^2$) of the fit to the bulk and the edge distributions. $R^2$ of the bulk decays for small values of $\kappa a^4$, since in this region, the occupation of the upper band is very low, hence the fit becomes less precise.

square fit of $f_{\mathbf{k}+}^{\mathrm{b}}$ to the Fermi-Dirac distribution. As expected, an effective temperature, grows with $\kappa$, as we increase the density of excitations. We also measured the coefficient of determination ($R^2$) for this fit (Fig. S4b). It shows an improvement of the fit as $\kappa$ increases. This is explained by the fact that for small $\kappa a^4$, the density of excitations is very small, which limits the precision of the fit. For higher $\kappa$ the distribution start to deviate from Fermi-Dirac form and the $R^2$ parameter decreases. A similar effect was found in Ref [S4].

### S5.2. Distribution of the edge

Here we estimate the scaling powers of the parameters $\delta$ and $T_\mathrm{e}$ of the quasi Fermi-Dirac distribution, $f_{\mathrm{QFD}}$, as function of $\kappa$. First, we observe that the majority of the excitations of the edge are concentrated near the center, the rest of the excitations are distributed approximately uniformly along the edge. We choose an arbitrary point $k_\mathrm{S}$, such that the majority of the excitations are within $0 < k < k_\mathrm{S}$. Then the density $n_\delta \equiv \int_{k_\mathrm{S}}^{k_R} \frac{dk_x}{2\pi} f_{k_x R}^\mathrm{e}$ is significantly smaller than $n_T \equiv n_\mathrm{e} - n_\delta \approx n_\mathrm{e}$. The parameters $\delta$ and $T_\mathrm{e}$ can be then expressed in terms of $n_\delta$ and $n_T$, plugging the approximation $f_{k_x R}^\mathrm{e} \approx f_{\mathrm{QFD}}$ in the definitions of $n_\delta$ and $n_T$, and assuming a constant edge velocity, $v_\mathrm{e}$, and $\delta \ll 1$. We obtain, $n_T \approx \frac{k_\mathrm{B} T_\mathrm{e} \ln(2)}{\pi a \Delta_1}$, and $n_\delta \approx \frac{k_R - k_\mathrm{S}}{4\pi} \delta$.

Now we are at the position to write the rate equation for $n_\delta$. The difference between this equation and Eq. (5b) in the main text is in an additional rate $R^{\mathrm{b} \to \mathrm{e}}$ that scatter particles from the lower band to the edge via photon mediated Umklapp processes. This rate is significant near the points where the edge states are close to the bulk. The rate equation for $n_\delta$ then reads

$$\dot{n}_\delta = R^{\mathrm{b} \to \mathrm{e}} + \gamma^{\mathrm{b} \to \mathrm{e}} n_\mathrm{b} - \Lambda^{\mathrm{e} \to \mathrm{b}} n_\mathrm{b} n_\delta - \gamma^{\mathrm{e} \to \mathrm{e}} n_\delta^2, \tag{S40}$$

with an approximate solution $n_\delta \approx \left( \frac{R^{\mathrm{b} \to \mathrm{e}} + \gamma^{\mathrm{b} \to \mathrm{e}} n_\mathrm{b}}{\gamma^{\mathrm{e} \to \mathrm{e}}} \right)^{\frac{1}{2}}$. The leading contribution to this expression is $\left( R^{\mathrm{b} \to \mathrm{e}} / \gamma^{\mathrm{e} \to \mathrm{e}} \right)^{\frac{1}{2}}$ which scales as $\kappa^{\frac{1}{2}}$.



\end{document}